\documentclass[11pt,a4paper]{article}

\usepackage{jcappub}
\usepackage{graphicx}
\usepackage[utf8]{inputenc}
\usepackage{listings}
\usepackage{enumerate}
\usepackage{amsmath}
\usepackage{mathtools}
\usepackage{amsfonts}
\numberwithin{equation}{section}
\usepackage{natbib}

\usepackage{xcolor}
\usepackage{physics,amssymb}
\usepackage{siunitx}
\usepackage{acronym}
\usepackage{xspace}
\usepackage{booktabs}

\hypersetup{colorlinks=true
,urlcolor=DARKBLUE
,anchorcolor=DARKBLUE
,citecolor=DARKBLUE
,filecolor=DARKBLUE
,linkcolor=DARKBLUE
,menucolor=DARKBLUE
,linktocpage=true
,pdfproducer=medialab
,pdfa=true
}

\newcommand{\bae}[1]{\begin{align} #1 \end{align}}

\newcommand{\bme}[1]{\begin{multline} #1 \end{multline}}

\definecolor{MONZA}{HTML}{CF000F}
\definecolor{DARKBLUE}{HTML}{00008b}
\definecolor{DARKMAGENTA}{HTML}{8b008b}
\definecolor{PURPLE}{HTML}{440099}

\acrodef{DW}{domain wall}
\acrodef{DE}{dark energy}
\acrodef{CMB}{cosmic microwave background}
\acrodef{SN}{supernova}
\newacroplural{SN}[SNe]{supernovae}
\acrodef{DESI}{Dark Energy Spectroscopic Instrument}
\acrodef{MCMC}{Markov Chain Monte Carlo}
\acrodef{ADM}{Arnowitt--Deser--Misner}
\acrodef{DESY5}{5-year data of Dark Energy Survey}

\newcommand{\mpl}{m_\mathrm{Pl}}
\newcommand{\DW}{\mathrm{DW}}
\newcommand{\m}{\mathrm{m}}
\newcommand{\obs}{\mathrm{obs}}
\newcommand{\cc}{\mathrm{c}}

\newcommand{\ue}{\mathrm{e}}
\newcommand{\ui}{\mathrm{i}}
\newcommand{\calL}{\mathcal{L}}
\newcommand{\um}{\mathrm{m}}

\newcommand{\calR}{\mathcal{R}}
\newcommand{\ur}{\mathrm{r}}
\newcommand{\bbZ}{\mathbb{Z}}

\title{Domain-wall Quintessence}
\author[a]{Nobufusa Kobayashi,}
\author[b,c,d,a]{Yuichiro Tada,}
\author[e, f]{Fuminobu Takahashi,}
\author[g]{and Takahiro Terada}

\affiliation[a]{Department of Physics, Nagoya University, \\
Furo-cho Chikusa-ku,
Nagoya 464-8602, Japan}
\affiliation[b]{Department of Applied Physics, University of Fukui, \\
Bunkyo 3-9-1, Fukui 910-8507, Japan}
\affiliation[c]{Department of Physics, Rikkyo University, \\ 
3-34-1 Nishi-Ikebukuro, Toshima, Tokyo 171-8501, Japan}
\affiliation[d]{Institute for Advanced Research, Nagoya University, \\
Furo-cho Chikusa-ku, 
Nagoya 464-8601, Japan}
\affiliation[e]{Department of Physics, Tohoku University, 
Sendai, Miyagi 980-8578, Japan} 
\affiliation[f]{Kavli IPMU (WPI), UTIAS, University of Tokyo, Kashiwa 277-8583, Japan}

\affiliation[g]{
Kobayashi-Maskawa Institute for the Origin of Particles and the Universe, Nagoya University, 
Furo-cho, Chikusa-ku, Nagoya 464-8602, Japan
}

\emailAdd{kobayasi.nobufusa.z1@s.mail.nagoya-u.ac.jp}
\emailAdd{ytada@u-fukui.ac.jp}
\emailAdd{fumi@tohoku.ac.jp}
\emailAdd{terada@eken.phys.nagoya-u.ac.jp}

\abstract{
We investigate a dark energy
model driven by a planar domain-wall-like structure with a thickness comparable to, or larger than, the current Hubble radius,
focusing on its intrinsic anisotropy and observational viability.
Near the centre
of the domain wall (DW), the spacetime is anisotropic, with distinct expansion rates parallel and perpendicular to the wall.
This anisotropic structure induces direction-dependent cosmic expansion and modifies photon geodesics from cosmological sources, leaving characteristic signatures in cosmological observables.
We confront the model with recent observational data.
We first compute the anisotropic Cosmic Microwave Background (CMB) temperature multipoles generated by the DW and impose constraints from the Planck 2018 measurements.
These constraints severely limit the allowed DW abundance, requiring the DW energy density
to be 
less than $\mathcal{O}(10^{-5})$ of the current critical density in order to suppress the
quadrupole
contributions.
We then perform a Markov Chain Monte Carlo (MCMC) analysis using Type Ia supernova (SNe Ia) data, including the Pantheon+~SH0ES and DESY5 samples, to compare the DW scenario with the standard $\Lambda$CDM model.
We find that although the DW naturally realises anisotropic accelerated expansion, the combined constraints from the CMB and SNe Ia favour the $\Lambda$CDM limit, in which the DW contribution is negligible, and the universe is effectively isotropic.
Our results demonstrate that a Hubble-scale domain wall is tightly constrained by current observations and can only play a subdominant role in the late-time cosmic acceleration.
}
\begin{document}
\noindent TU-1308

\maketitle
\flushbottom

\section{Introduction}
Observations have revealed that the universe is currently undergoing accelerated expansion, which is commonly attributed to dark energy (DE). The cosmological constant is the simplest candidate for DE, 
as adopted in the current standard cosmological model, the $\Lambda$CDM model. 
The $\Lambda$CDM model is based on the cosmological principle, which assumes that the universe is homogeneous and isotropic on large scales. 

Recently, the  \ac{DESI} has shown the possibility that DE is time-dependent~\cite{DESI:2024mwx, DESI:2024hhd, DESI:2025zgx, DESI:2025fii}. If confirmed, this would point to a deviation from the $\Lambda$CDM model and require a modification or extension of the standard cosmological framework. 
Various models have been proposed to explain dynamical \ac{DE}, including quintessence models~\cite{Tsujikawa:2013fta,Copeland:2006wr}, in which the potential energy of scalar fields drives the accelerated expansion. 
Cosmic inhomogeneities have also been studied as a possible explanation of the \ac{DESI} results~\cite{Ginat:2026fpo,Alnes:2006uk,Alnes:2006pf,Enqvist:2006cg}. Some quintessence models for explaining dynamical \ac{DE} can induce large inhomogeneities~\cite{Yoshioka:2026qyd}, while other models~\cite{Khoury:2025txd} can produce topological defects, such as domain walls, in certain regions of parameter space. 

Recent high-precision observations have shown hints that the universe may not be perfectly isotropic even on the largest scales. The most prominent example of large-scale anisotropy is found in the \ac{CMB} multipoles~\cite{Planck:2018nkj}. It is well established that the CMB dipole, which corresponds to $10^{-4}\text{--}10^{-3}$ temperature fluctuations compared to the usual $10^{-5}$ fluctuations, can be attributed to the peculiar motion of the observer. In an isotropic universe, the CMB dipole can therefore be treated as a purely kinematical effect. However, it remains possible that this anisotropy arises from other physics. In fact, observations by WMAP and Planck have revealed several anisotropic anomalies that may challenge the prevailing $\Lambda$CDM model and the cosmological principle. For example, the alignment of the low multipole moments has attracted considerable attention~\cite{10.1093/mnras/staf461,Aluri:2023xmb,Amo:2025ghl}. It has been known that the quadrupole and octopole components are unnaturally oriented relative to the ecliptic plane~\cite{CosmoVerseNetwork:2025alb}.
It is also suggested that the recent accelerated expansion may be anisotropic, and that the preferred direction is close to the CMB dipole direction~\cite{Krishnan:2021jmh}. In addition to the expansion rate~\cite{Bahr-Kalus:2012yjc,Cai:2013lja,Sah:2024csa,Verma:2023huz,Verma:2024lex}, other anisotropic features have also been reported, such as large-scale bulk flows and quasar polarisation vectors~\cite{CosmoVerseNetwork:2025alb}.

Models based on topological defects provide a natural class of quintessence scenarios with spatial anisotropy. Topological defects are topologically stable structures in which the field configuration is spatially inhomogeneous. If their characteristic size is comparable to, or larger than, the current Hubble radius, they can drive the accelerated expansion of the universe. Topological inflation and the associated eternal inflation, which are based on this idea in the early universe, have been extensively studied~\cite{Sakai:1995nh,Vilenkin:1994pv}. Similarly, this mechanism can be applied to dark energy: quintessence models based on topological defects that are sufficiently stable to survive until the present epoch and behave as dark energy have been proposed as candidates for the current accelerated expansion of the universe, and are constrained by observations~\cite{Sousa:2015cqa,BuenoSanchez:2011wr,Grande:2011hm,Perivolaropoulos:2012mca,An:2025vfz,Battye:1999eq,Friedland:2002qs,Lee:2025yvn}.

The purpose of this paper is to investigate whether such an anisotropic DE component, realised by a Hubble-scale domain wall (\ac{DW}), namely, a domain-wall structure whose thickness is comparable to, or larger than, the current Hubble radius, can be consistent with current cosmological observations. Hereafter, we refer to such a configuration as a Hubble-scale DW.
In particular, we emphasise that the central region of the Hubble-scale \ac{DW} is intrinsically anisotropic, 
leading to direction-dependent expansion rates.
This is in sharp contrast to monopole-based quintessence models, in which an observer located at the centre can see an isotropic configuration owing to the spherical symmetry~\cite{BuenoSanchez:2011wr}. For a planar \ac{DW}, by contrast, the direction normal to the wall is distinguished from the direction parallel to it even near the wall centre, so that anisotropic cosmic acceleration is unavoidable.
We confront this model with observational data in two complementary ways.
First, we derive the anisotropic CMB temperature multipoles induced by the DW geometry and impose constraints from the Planck 2018 observations.
Second, we perform a \ac{MCMC} analysis using \acp{SN} Ia data, including the Pantheon+~SH0ES (Supernovae and $H_0$ for the Equation of State of dark energy)~\cite{Riess:2021jrx,Brout:2022vxf} compilation
and the \ac{DESY5}~\cite{DES:2025sig}, to compare the DW scenario with the standard $\Lambda$CDM model.
By combining these probes, we demonstrate that the allowed DW energy fraction is extremely small and that current data strongly favour the $\Lambda$CDM limit.

The rest of this paper is organised as follows.
In section~\ref{dynamics}, we study the dynamics of a cosmological model with a planar DW and clarify its anisotropic properties near the DW centre.
In section~\ref{Sec_geodesic}, we derive photon geodesics. In section~\ref{constraintCMB}, we present the corresponding observational constraints from CMB anisotropies, and in section~\ref{MCMC}, we perform \ac{MCMC} parameter estimation using 
\ac{SN} distance moduli.
Finally, section~\ref{conclusion} is devoted to conclusions and discussion. Appendix~\ref{near} provides approximate solutions in both the near- and far-\ac{DW} regions.
In appendix~\ref{higher}, we discuss the constraints from higher \ac{CMB} multipoles.
Appendix~\ref{jointed} presents the \ac{MCMC} parameter estimation based on multiple observational datasets.

\section{Dynamics of the planar DW universe}\label{dynamics}

Throughout this paper, we consider a single planar \ac{DW}-like structure with a thickness comparable to or greater than the present Hubble radius. Here we refer to it as a ``\ac{DW}-like structure'' because it extends over super-Hubble scales and does not represent either a conventional domain wall in the scaling regime or the usual energy-minimising domain-wall solution subject to asymptotic boundary conditions.
For simplicity, however, we hereafter refer to this configuration simply as a \ac{DW}.
We model it by a planar toy setup. Although idealised, this approximation should be reasonable on scales smaller than the wall thickness, since smaller-scale structure is energetically disfavoured and domain walls generally tend to straighten.

\subsection{Setup}\label{setup}

We consider the following action:
\begin{equation}\label{Action}
    S=\int\dd[4]{x}\sqrt{-g}\left[\frac{\mpl^2}{16\pi}(\mathcal{R}-2\Lambda)-\frac{1}{2}(\partial_\mu\Phi)^2-V(\Phi)+\mathcal{L}_\um\right].
\end{equation}
Here, 
$\mpl$ is the Planck mass, and $\calR$ is the Ricci curvature of the spacetime metric.
We also include a 
cosmological constant, denoted by $\Lambda$.
$\Phi$ is a real scalar field, and we assume a $\bbZ_2$-symmetric double-well potential of the form:
\bae{
    V(\Phi)=\frac{1}{4}\lambda(\Phi^2-\eta^2)^2,
}
where $\eta$ is the vacuum expectation value, and $\lambda$ is the coupling constant. 
For simplicity, we consider an infinite planar \ac{DW} configuration. The plane on which the \ac{DW} lies  can be taken to be the $yz$ plane in Cartesian coordinates, without loss of generality. That is, $\Phi$ is a function only of $x$ and $t$:
\begin{equation}
\label{config}
    \Phi=\Phi(x,t),
\end{equation}
and satisfies the boundary conditions
\begin{equation}
\label{boundary}
    \Phi(0,t)=0
    \qc \Phi(\pm\infty,t)=\pm\eta.
\end{equation}
The spacetime metric that is  
homogenous only in the $y$- and $z$- directions is given by 
\begin{equation}
\label{metric}
    \dd{s^2}=-\dd{t^2}+A^2(x,t)\dd{x^2}+B^2(x,t)(\dd{y^2}+\dd{z^2}),
\end{equation}
where the two scale factors $A(x,t)$ and $B(x,t)$ are functions of $x$ and $t$.
$\calL_\um$ represents the Lagrangian density for the matter fluids.

Since we are considering the late-time universe, we assume that the energy-momentum tensor $T_{\mu\nu}$ is dominated by matter, \ac{DW}, and the cosmological constant:
\begin{equation}
\label{tensor}
    T_{\mu\nu}=    T_{\mu\nu}^{(\m)}+T_{\mu\nu}^{(\DW)}+T_{\mu\nu}^{(\Lambda)}.
\end{equation}
Assuming that the matter is a perfect fluid, its energy-momentum tensor is given by
\begin{equation}\label{tensormat}
    T_{\mu\nu}^{(\m)}=\rho_\m u_\mu u_\nu.
\end{equation}
Here, $\rho_\m=\rho_\m(x,t)$ is the matter density, and $u^\mu$ is the 4-velocity of the perfect fluid, which can be written as
\begin{equation}
    u^\mu=\left(\frac{1}{\sqrt
    {1-v^2}},\frac{v}{A\sqrt
    {1-v^2}},0,0\right).
    \label{4velo}
\end{equation}
where $v=v(x,t)$ is the velocity of the perfect fluid in the $x$-direction.
The energy-momentum tensor for the \ac{DW} is given by
\begin{align}
    T_{\mu\nu}^{(\DW)}
    =\partial_\mu\Phi\partial_\nu\Phi-g_{\mu\nu}\left[\frac{1}{2}\left(-\dot{\Phi}^2+\frac{\Phi'^2}{A^2}\right)+\frac{1}{4}\lambda(\Phi^2-\eta^2)^2\right].
    \label{tensordom}
\end{align}
Here, the prime and dot denote partial derivatives with respect to $x$ and $t$, respectively.   
The \ac{DW} energy density reads
\begin{align}
    \rho_\DW=T^{(
    \DW)}_{00}=\frac{1}{2}\dot{\Phi}^2+\frac{\Phi'^2}{2A^2}+\frac{\lambda}{4}(\Phi^2-\eta^2)^2.
    \label{domenergy}
\end{align}
On the other hand, the energy-momentum tensor for the cosmological constant is given by
\bae{
    T_{\mu\nu}^{(\Lambda)}=-\frac{\mpl^2}{8\pi}\Lambda g_{\mu\nu},
}
which leads to its constant energy density,
\bae{
    \rho_\Lambda=T_{00}^{(\Lambda)}=\frac{\mpl^2}{8\pi}\Lambda.
}

The relevant equations of motion for the metric and the scalar field are the Einstein and the Klein--Gordon equations:
\begin{align}
\label{ham}
    -G_0{}^0&=K_2{}^2(2K-3K_2{}^2)-\frac{2B''}{A^2B}-\frac{B'^2}{A^2B^2}+\frac{2A'B'}{A^3B}\notag\\
    =-\frac{8\pi}{\mpl^2}T_0{}^0 &=\frac{8\pi}{\mpl^2}\left[\frac{\dot{\Phi}^2}{2}+\frac{{\Phi'}^2}{2A^2}+\frac{\lambda}{4}(\Phi^2-\eta^2)^2+\frac{\rho_\m}{1-v^2}+\rho_\Lambda\right]
    ,
\end{align}
\begin{align}
\label{mom}
    \frac{1}{2}G_{01}&=({K_2{}^2})'+\frac{B'}{B}(3K_2{}^2-K)\\
   =\frac{8\pi}{\mpl^2}\left(\frac{1}{2}T_{01}\right)
   &=\frac{4\pi}{\mpl^2}\left[\dot{\Phi}{\Phi}'-\frac{v}{1-v^2}A\rho_\m\right]
    ,
\end{align}
\begin{align}
\label{sum}
    \frac{1}{2}(G_1{}^1+G_2{}^2+G_3{}^3-G_0{}^0)&=\dot{K}-(K_1{}^1)^2-2(K_2{}^2)^2\notag\\
   =\frac{8\pi}{\mpl^2}\left[\frac{1}{2}(T_1{}^1+T_2{}^2+T_3{}^3-T_0{}^0)\right] &=\frac{8\pi}{\mpl^2}\left[\dot{\Phi}^2-\frac{\lambda}{4}(\Phi^2-\eta^2)^2+\frac{1}{2}\frac{1+v^2}{1-v^2}\rho_\m-\rho_\Lambda\right]
    ,
\end{align}
\begin{align}
\label{odd}
    -\calR_2{}^2-G_0{}^0&=\dot{K}_2{}^2+\frac{B'^2}{2A^2B^2}-\frac{3}{2}(K_2{}^2)^2\notag\\
    =\frac{8\pi}{\mpl^2}\left(-\frac{T_0{}^0}{2}+\frac{T_1{}^1}{2}\right)
    &=\frac{4\pi}{\mpl^2}\left[\frac{\dot{\Phi}^2}{2}+\frac{{\Phi'}^2}{2A^2}-\frac{\lambda}{4}(\Phi^2-\eta^2)^2+\frac{v^2}{1-v^2}\rho_\m-\rho_\Lambda\right]
    ,
\end{align}
and
\begin{equation}
\label{EL}
    \ddot{\Phi}-K\dot{\Phi}-\frac{{{\Phi}}''}{A^2}-\left(-\frac{A'}{A}+\frac{2B'}{B}\right)\frac{\Phi'}{A^2}+\lambda\Phi(\Phi^2-\eta^2)=0
    ,
\end{equation}
where $\calR_{\mu\nu}$ and $G_{\mu\nu}=\calR_{\mu\nu}-\frac{1}{2}g_{\mu\nu}\calR$ are the Ricci and Einstein tensors, respectively. In the above equations, we introduce the extrinsic curvature $K_{ij}$, which is expressed in terms of the \ac{ADM} formalism as follows~\cite{Misner:1973prb}: 
\begin{align}
    K_{ij}=\frac{1}{2\alpha}(-\partial_0\gamma_{ij}+D_i\beta_j+D_j\beta_i),
\end{align} 
where $\gamma_{ij}$ is the 3-dimensional metric on the spatial hypersurface, $\alpha$ and $\beta_i$ are the lapse function and shift vector, respectively, and $D_i$ is the covariant derivative compatible with $\gamma_{ij}$. The non-vanishing components of this extrinsic curvature in the equations of motion are given by
\begin{equation}
    K_1{}^1
    =-\frac{\dot{A}}{A}\qquad \text{and} \qquad K_2{}^2=K_3{}^3
    =-\frac{\dot{B}}{B},
    \label{Ks}
\end{equation}
and $K \equiv K_i{}^i$ is the trace of the extrinsic curvature. The spatial indices are raised and lowered by $\gamma_{ij}$ and its inverse $\gamma^{ij}$. 
We do not explicitly impose the boundary conditions~\eqref{boundary}, but monitor that the numerical solution satisfies them with appropriate initial conditions.\footnote{As explained in Ref.~\cite{Sakai:1995nh}, eq.~\eqref{odd} can be used to solve for the time evolution of $K_2{}^2(0, t)$ (only at $x = 0$), and then eq.~\eqref{mom} can be used to solve for $K_2{}^2(x, t)$ along the $x$-direction for each time step. In this sense, we can regard eq.~\eqref{odd} as playing the role of the boundary condition, although the algorithm used by the \texttt{NDSolve} function in \textit{Wolfram Mathematica} in our codes may not be the same. Time evolution of other quantities, $A$, $B$, $K$, and $\Phi$ can be obtained by solving for eqs.~\eqref{Ks}, \eqref{sum}, and \eqref{EL}, respectively. 
}
The energy-momentum conservation 
for the matter fluid leads to its evolution equations
\bae{
    \dot{v}&=v(v^2-1)\frac{\dot{A}}{A}-\frac{vv'}{A}
    ,
    \label{conserve1} \\
    \label{conserve2}
    \frac{\dot{\rho}_\m}{\rho_\m}&=(v^2-1)\frac{\dot{A}}{A}-\frac{v'}{A}-2\frac{\dot{B}}{B}-\frac{v}{A}\frac{\rho'}\rho{-2\frac{v}{A}\frac{B'}{B}}.
}

Let us introduce parameters that quantify the averaged spatial expansion and the anisotropy of expansion. 
First, the effective scale factor is defined by
\begin{align}
    \bar{a}(x,t) \equiv \left[ A(x,t) B^2(x,t) \right]^{1/3},
     \label{abar}
\end{align}
 and the Hubble parameter associated with $\bar{a}$ is 
\begin{align}
     H_{\bar{a}} = \frac{\dot{\bar{a}}}{\bar{a}} = -\frac{K}{3}.
     \label{newhubble}
\end{align}
 The anisotropy $\sigma$ is defined by
\begin{align}
 \sigma \equiv \frac{-K_1{}^1 - H_{\bar{a}}}{H_{\bar{a}}} = \frac{2(-K_1{}^1 + K_2{}^2)}{-K_1{}^1 - 2K_2{}^2}.\label{sigma}
\end{align}
Note that these are locally defined quantities that depend on $x$. 

In the Minkowski limit, $A \approx B \approx 1$, and assuming $\rho_\m=0$, eq.~\eqref{EL} leads to the static DW solution:
\bae{
    \label{tanh}
    \Phi \approx \eta \tanh{\left(\frac{
    x}{\sqrt{2}D}\right)},
    }
where
\bae{
    \label{thickness}
    D 
    \coloneqq\lambda^{-1/2} \eta^{-1}
    ,
}
is the characteristic thickness of the wall.
Note that $\Phi'$ does not vanish near the central plane of the wall, $\abs{x}\ll \sqrt{2}D$.
This gives rise to the difference between the elements of the energy-momentum tensor,
\begin{align}
\label{T11}
    T_1{}^1 &
    =\frac{1}{2}\dot{\Phi}^2 + \frac{\Phi'^2}{2A^2} - \frac{\lambda}{4}(\Phi^2 - \eta^2)^2, \\
\label{T22}
    T_2{}^2 = T_3{}^3 &= \frac{1}{2}\dot{\Phi}^2 - \frac{\Phi'^2}{2A^2} - \frac{\lambda}{4}(\Phi^2 - \eta^2)^2 .
\end{align}
Supposing that $\Phi$ is quasi-static, their Taylor expansion  
around $x = 0$ reads 
\begin{align}
    T^1_1 &\approx \frac{\lambda\eta^4}{4}(A^{-2}-1) + \mathcal{O}(x^2), \\
    T^2_2 = T^3_3 &\approx -\frac{\lambda\eta^4}{4}(A^{-2}+1) + \mathcal{O}(x^2).
\end{align}
That is, $T_1{}^1$ is different from $T_2{}^2=T_3{}^3$ (i.e., anisotropic) even at the wall centre, already at the ${\cal O}(x^0)$ level.

The quasi-static or slow-roll condition of $\Phi$ is obtained as follows.
Suppose that $\Phi$ is initially quasi-static and linear in $x$ around the centre,
\bae{
    \ddot{\Phi} \simeq 0,\quad \Phi'' \simeq 0, 
}
and we take the initial time at the deep matter-dominated era.  Then the initial scale factors $A_\ui$ and $B_\ui$ are almost homogeneous because the energy density of \ac{DW} is negligible at that epoch,
\bae{
    \partial_xA_\ui\approx\partial_xB_\ui\approx0.
}
Thus, we obtain
\begin{align}
\Phi(x,t) \approx \Phi(x,t_0) \exp\!\left(-\frac{\lambda\eta^2}{K}(t - t_0)\right),
\label{exp}
\end{align}
where $t_0$ is the present time.  
Since we are interested in the late-time evolution,
the relevant time interval $t - t_0$ can be supposed to be of order the Hubble time $\sim(-G_0^{\,0})^{-1/2}\approx1/\sqrt{K_2{}^2(2K-3K_2{}^2)}$, where we neglect the spatial derivatives.
Therefore, the field $\Phi$ remains effectively static on cosmological timescales if the 
absolute value of the exponent is
\begin{align}
\label{slowroll}
   \abs{\frac{\lambda\eta^2}{K}(t-t_0)}\approx\left|\frac{\lambda\eta^2}{K\sqrt{K_2{}^2(2K - 3K_2{}^2)}}\right|_{t = t_0} \ll 1.
\end{align}
See appendix~\ref{near} for more detailed analytical solutions.

\subsection{Numerical solutions}\label{numerical}

We resort to a numerical method to
solve the dynamics of the planar \ac{DW} universe. To this end, we use the following dimensionless variables:
\begin{align}
    \bar{x} \equiv \sqrt{\lambda}\eta x, \quad \bar{t} \equiv \sqrt{\lambda}\eta t, \quad \bar{\Phi} \equiv \frac{\Phi}{\eta},
    \quad
    \bar{\rho}_i\equiv \frac{\rho_i}{\lambda\eta^4}, \quad \text{and} \quad \bar{\eta} \equiv \frac{\eta}{\mpl},
\end{align}
where $i= \m$, $\DW$, or $\Lambda$. 
With this normalisation, 
eqs.~(\ref{ham})--(\ref{EL}), and (\ref{Ks})--(\ref{sigma}) are transformed as follows: all dimensionful variables are replaced with their dimensionless counterparts, denoted by bars, and parameters $\lambda$ and $\eta$ are effectively set to unity.
The coefficient $8\pi/{\mpl^2}$ appearing in the right-hand side of eqs.~(\ref{ham})--(\ref{odd}) is then replaced with $8\pi\bar{\eta}^2$.
Using this normalisation, we solve the set of eqs.~(\ref{sum})--(\ref{odd}) and monitor the accuracy of the obtained solutions
by the Hamiltonian constraint~\eqref{ham} and
the momentum constraint~\eqref{mom}.
We have confirmed that these constraints are satisfied to an accuracy better than 1\% in our calculation.

The dimensionless initial time $\bar{t}_\ui$ of the system is 
chosen in the matter-dominated era, when the matter distribution is homogeneous and isotropic. 
In section~\ref{constraintCMB}, we consider both the cosmological-constant-dominated late-time universe and the \ac{DW}-dominated late-time universe. In the former case, we define the initial time as follows:
\begin{align}
\frac{\bar{\rho}_\mathrm{m}(\bar{x}, \bar{t}_\ui)}{\bar{\rho}_\Lambda} = 18.
\label{initime}
\end{align}
In the \ac{DW}-dominated case, we define the initial time by replacing $\bar{\rho}_\Lambda$ with $V(\Phi(\bar{x}=0, \bar{t}_\ui))$.

For the initial field profile of $\Phi$, we adopt the static \ac{DW} solution~(\ref{tanh}).
These assumptions are summarised as the following set of initial conditions:
\begin{align}
    \bar{v}(\bar{x},\bar{t}_\ui) &= 0,\quad\bar{t}_\ui = \sqrt{\frac{1}{6 \bar{\rho}_\m(\bar{x}, \bar{t}_\ui)\pi \bar{\eta}^2}},
    \\
    A(\bar{x}, \bar{t}_\ui) &= B(\bar{x}, \bar{t}_\ui) = 1, \quad\bar{\Phi}(\bar{x}, \bar{t}_\ui) = \tanh{\left(\frac{\bar{x}}{\sqrt{2}}\right)},\quad\dot{\bar{\Phi}} = 0, \label{eq: Ai Bi Phii PIi}
    \\
    \bar{K}_2{}^2(\bar{x}, \bar{t}_\ui) &= -\frac{\sqrt{8\pi}\bar{\eta}}{\sqrt{3}} \sqrt{ \bar{\rho}_\m(\bar{x}_{\max}, \bar{t}_\ui) + \bar{\rho}_{\DW}(\bar{x}_{\max}, \bar{t}_\ui)+\bar{\rho}_\Lambda}, \label{icon1} \\
    \bar{K}(\bar{x}, \bar{t}_\ui) &= \frac{1}{2} \left[ \frac{8\pi \bar{\eta}^2}{\bar{K}_2{}^2(\bar{x}, \bar{t}_\ui)}\left(\frac{1}{2\cosh^4(\bar{x}/\sqrt{2})} + \bar{\rho}_\m(\bar{x}, \bar{t}_\ui)+\bar{\rho}_\Lambda \right) + 3\bar{K}_2{}^2(\bar{x}, \bar{t}_\ui) \right] .\label{icon2}
\end{align}
Substituting eq.~(\ref{initime})--(\ref{eq: Ai Bi Phii PIi}) into eq.~(\ref{mom}), we obtain $(K_2{}^2)'(x,t_\ui)=0$.
This implies that $K_2{}^2$ has no spatial dependence.
Therefore, without loss of generality, we evaluate the quantity in the limit $x\gg1$.
Far away from the DW, the spacetime becomes homogeneous and isotropic.
In this region, we can use the relation $K = 3K_2{}^2$.
Substituting this relation into eq.~(\ref{ham}), we obtain eq.~(\ref{icon1}), where $\bar{x}_{\max}\gg1$; we choose $\bar{x}_{\max} = 7$ for our analysis.
Since the spacetime is homogeneous and isotropic at the initial time, the scalar field satisfies eq.~(\ref{eq: Ai Bi Phii PIi}).
Substituting these conditions into eq.~(\ref{ham}), we finally obtain eq.~(\ref{icon2}).

These initial conditions correspond to a single planar \ac{DW}-like configuration
whose characteristic scale is comparable to, or larger than, the Hubble radius
at the matter-dominated era, where the matter distribution is homogeneous and isotropic.
In the observational analysis below, we treat the observer's distance from the \ac{DW} centre as a free parameter, so that the \ac{DW} centre may lie either inside or outside our Hubble patch.
We discuss the resulting CMB anisotropy in section~\ref{constraintCMB}.

The present time $t_0$ is determined by the condition,
\begin{align}
    \frac{\rho_\m(0, t_0)}{V(\Phi(x=0,t_0))+\rho_\Lambda} = \frac{1-\tilde{\Omega}_{\rm{DW}}-\Omega_{\Lambda}}{\tilde{\Omega}_{\rm{DW}}+\Omega_{\Lambda}},
\end{align}
where the current density parameters $\tilde{\Omega}_\DW \equiv V(\Phi(x = 0, t_0))/\rho_\text{crit}$ and $\Omega_\Lambda \equiv \rho_\Lambda / \rho_\text{crit}$ are fixed as input parameters, where $\rho_\text{crit}$ is the present critical energy density of the universe.
Note that $\tilde{\Omega}_{\rm{DW}}$ denotes the present energy density parameter associated only with the potential energy of the \ac{DW} at its central plane, rather than the total energy density of \ac{DW}.
Hereafter, we mainly fix $\Omega_\Lambda=0.7$, though we also consider the $\Omega_\Lambda=0$ case at the end of section~\ref{constraintCMB}.

Therefore, the remaining model parameters are $(\eta, \tilde{\Omega}_{\DW}, h)$, where the normalised Hubble constant $h=H_0/(\SI{100}{km/s/Mpc})$
effectively sets the overall scale of the system.
We will
use the physical thickness of the DW, $D$,  as an input model parameter instead of
$\eta$.
Note that the coupling constant $\lambda$ is determined indirectly from the following relation:
\begin{align}
   V(\Phi(x=0,t_0))=\frac{\lambda \eta^4}{4} &= \frac{3\mpl^2}{8\pi}\tilde{\Omega}_{
    \DW}
    H_0^2.
\end{align}
The physical dimension of the length and time is recovered as follows:
\begin{align}
    x &= \frac{\bar{x}}{\lambda^{1/2} \eta} \approx 4.37
    \frac{\bar{\eta}}{\tilde{\Omega}_\DW^{1/2}h}\bar{x}\,\si{Gpc}\label{dimx},\\
    t &=\frac{\bar{t}}{\lambda^{1/2} \eta}\approx 14.3 
    \frac{\bar{\eta}}{\tilde{\Omega}_\DW^{1/2}h}\bar{t}\,\si{Gyr}.\label{dimt}
\end{align}
Equation~\eqref{eq: Ai Bi Phii PIi} indicates that the dimensionless thickness of \ac{DW} is always unity. 

\begin{figure}
    \centering
    \begin{tabular}{c}
    \begin{minipage}{0.45\columnwidth}
        \centering
        \includegraphics[width=0.95\linewidth]{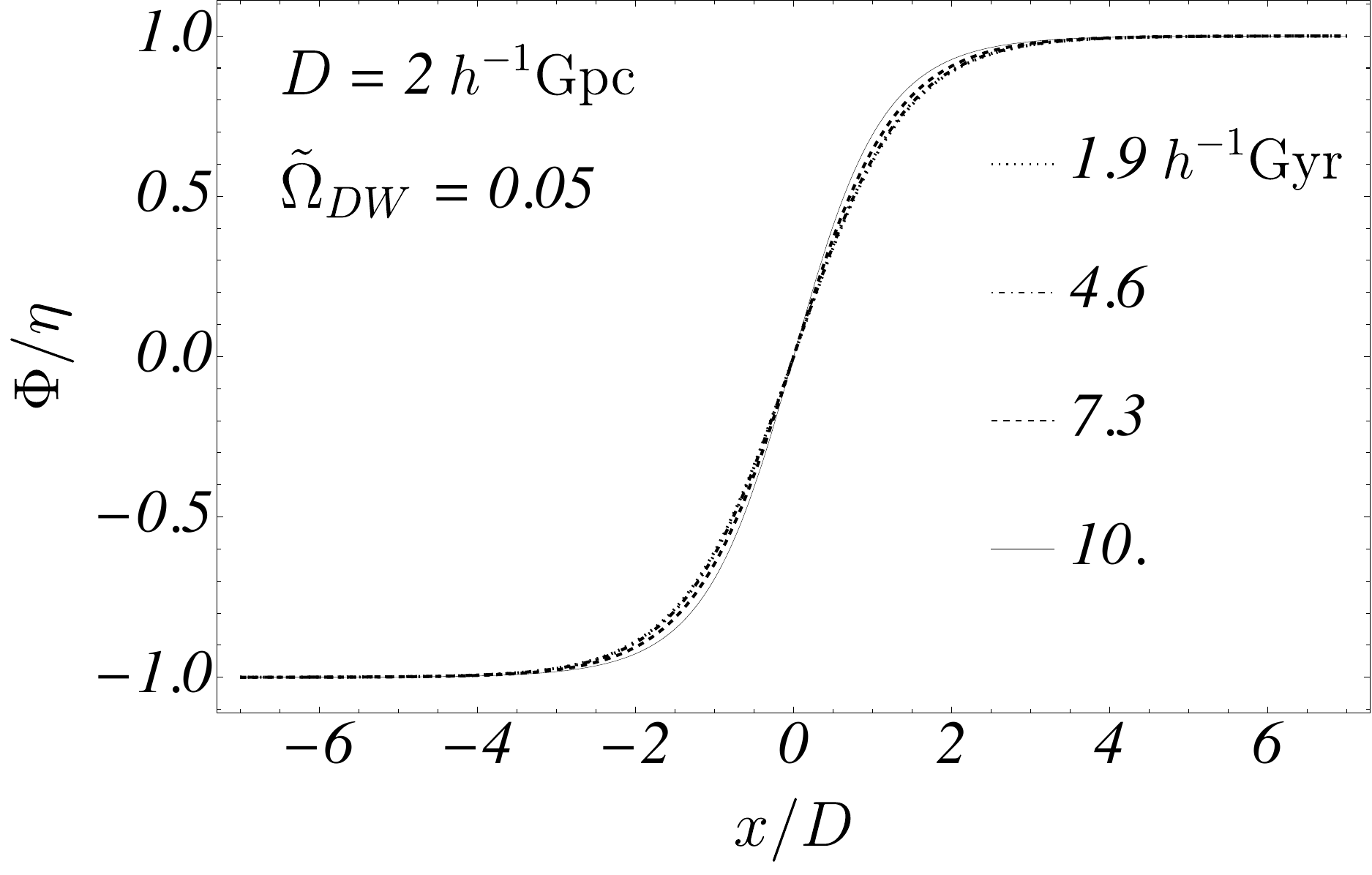}
    \end{minipage}
    \begin{minipage}{0.45\columnwidth}
        \centering
        \includegraphics[width=0.95\linewidth]{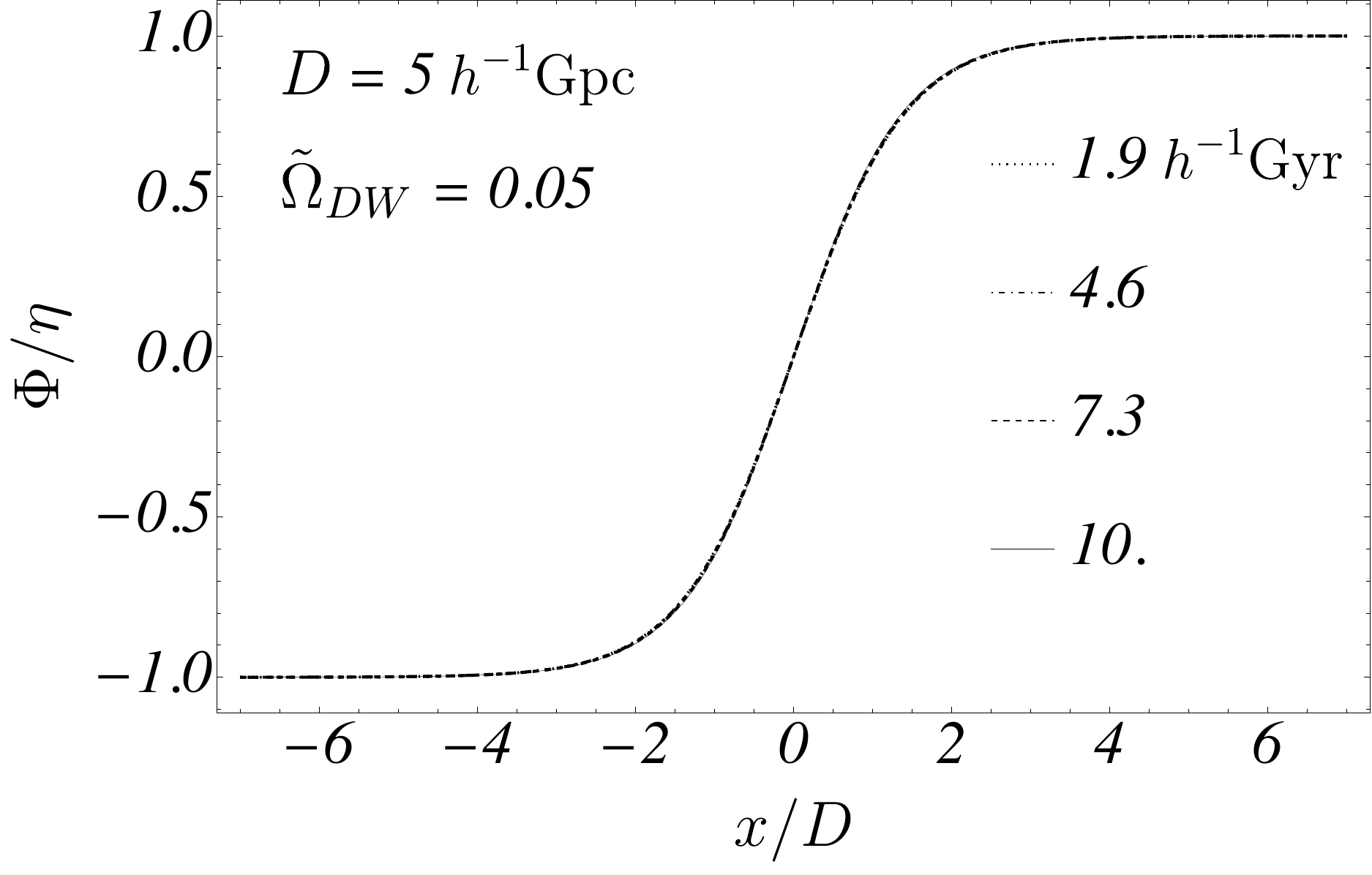}
    \end{minipage}
    \end{tabular}
    \caption{Time evolution of the field profile
    for $(D, \tilde{\Omega}_{\DW}) = (2h^{-1}\,\si{Gpc}, 0.05)$ (left)
    and $(D, \tilde{\Omega}_{\DW}) = (5h^{-1}\,\si{Gpc}, 0.05)$ (right). The line styles represent different times: dotted ($t_\ui$), dot-dashed ($(2t_\ui + t_0)/3$), dashed ($(t_\ui + 2t_0)/3$), and solid ($t_0$). A thicker DW evolves more slowly with time.
    }
    \label{fig:phi}
\end{figure}

Figure~\ref{fig:phi} shows the evolution of the scalar field profile $\Phi(x, t)$ from the initial time $t_\ui$ to the present time $t_0$ for 
$(D,\tilde{\Omega}_\DW)=(2h^{-1}\,\si{Gpc},0.05)$ and $(D,\tilde{\Omega}_\DW)=(5h^{-1}\,\si{Gpc},0.05)$.
For $D=2h^{-1}\,\si{Gpc}$, the slow-roll condition (\ref{slowroll}) is not satisfied, and the field 
gradually evolves toward the minimum of the potential over a cosmological timescale.
In contrast, for 
$D = 5h^{-1}\,\si{Gpc}$, the condition is satisfied, and the field remains relatively unchanged.
In both cases, the profile well satisfies the boundary conditions~\eqref{boundary}.

\begin{figure}
    \centering
    \begin{tabular}{c}
        \begin{minipage}{0.95\hsize}
            \centering
            \includegraphics[width=0.95\linewidth]{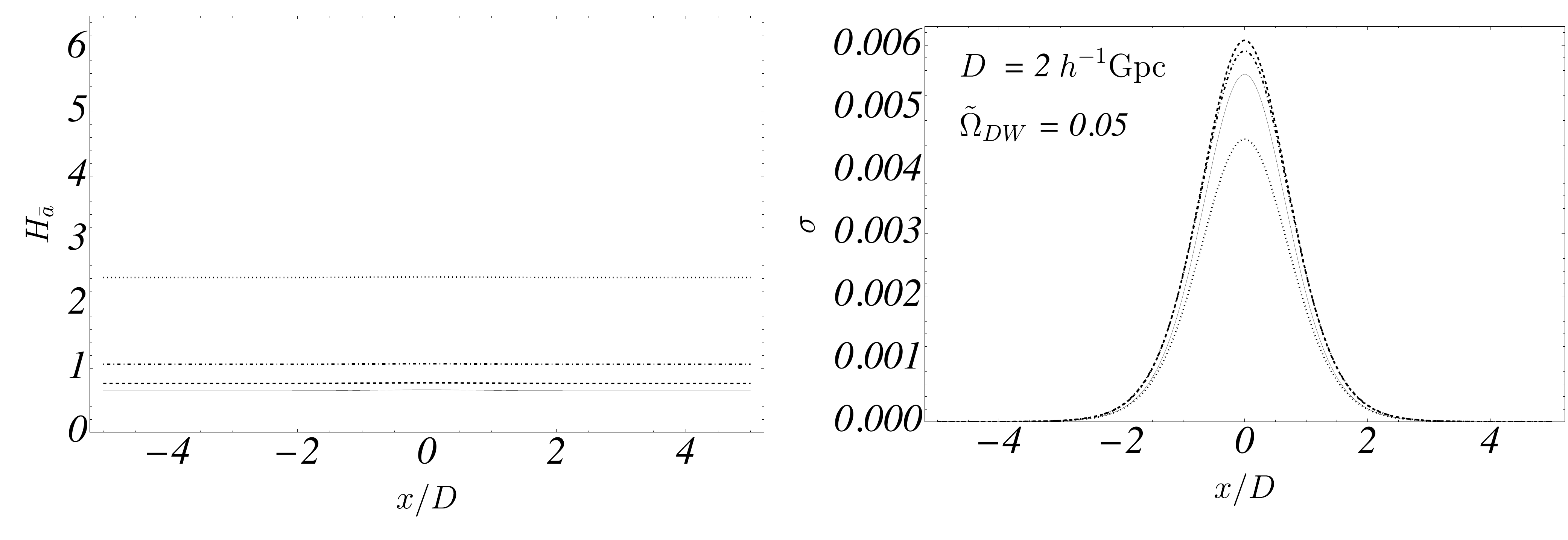}
        \end{minipage} \\
        \begin{minipage}{0.95\hsize}
            \centering
            \includegraphics[width=0.95\linewidth]{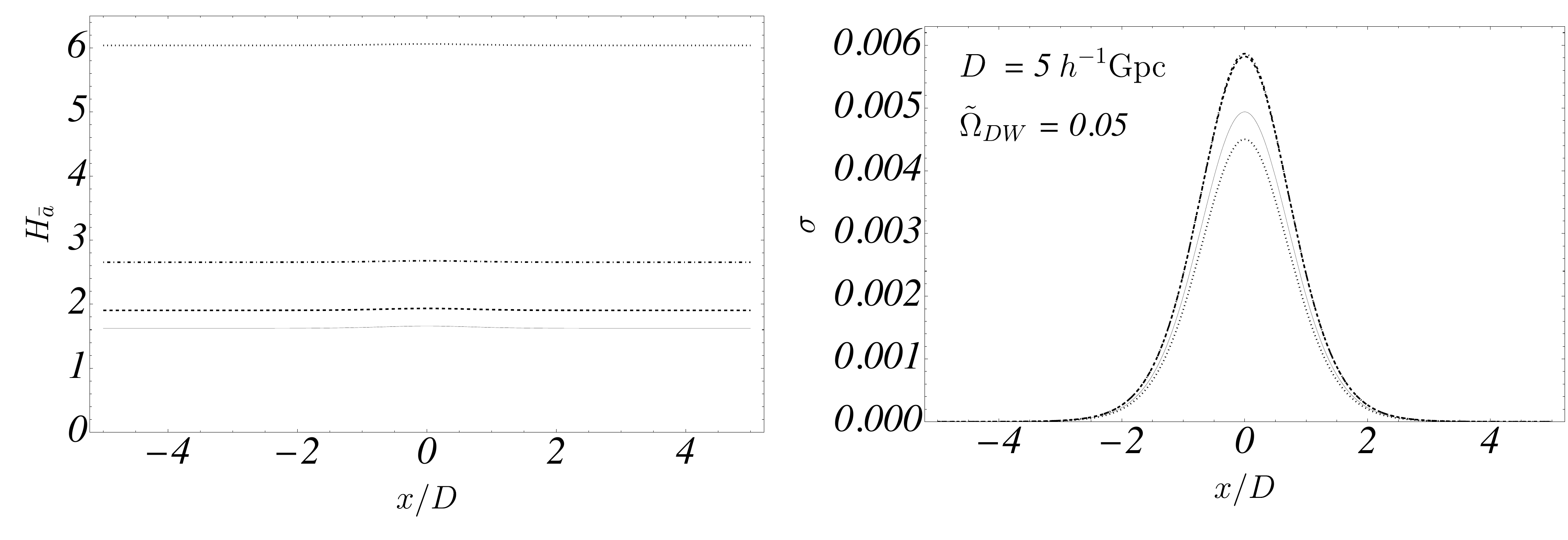}
        \end{minipage}
    \end{tabular}
    \caption{Time evolution of 
    the averaged Hubble parameter $H_{\bar{a}}$ (left) and the anisotropy parameter $\sigma$ (right) for $(D,~\tilde{\Omega}_{\DW})=(2~h^{-1}\si{Gpc}, 0.05)$ (top) and $(D,~\tilde{\Omega}_{\DW})=(5~h^{-1}\si{Gpc}, 0.05)$ (bottom). The line styles are the same as figure~\ref{fig:phi}.
    $H_{\bar{a}}$ is slightly larger near the wall than that in the distant region. The condition $\sigma > 0$ near the wall indicates an enhanced expansion rate perpendicular to the wall. The anisotropy initially grows over time, but decreases once the \ac{DW} starts to
    dominate the matter fluid, similar to the behaviour observed in topological inflation.}
    \label{AB}
\end{figure}

Figure~\ref{AB} shows the time evolution of the mean Hubble parameter $H_{\bar{a}}(x, t)$ and the anisotropy $\sigma(x, t)$. We see that 
the spacetime is
isotropic at large distances, but anisotropic ($\sigma\neq0$) near the wall as discussed in section~\ref{setup}. 
In particular, $\sigma$ is positive near the wall, which indicates that the expansion rate in the direction perpendicular to the wall is enhanced near the wall compared with that in distant region. The anisotropy initially grows over time; however, as the \ac{DW} grows, the anisotropy is reduced, similar to the behaviour in the topological inflation. We check that for smaller $\tilde{\Omega}_{\DW}$, the inhomogeneity and anisotropy become smaller.

\begin{figure}
    \centering
    \begin{tabular}{c}
        \begin{minipage}{0.95\hsize}
            \centering
            \includegraphics[width=0.95\linewidth]{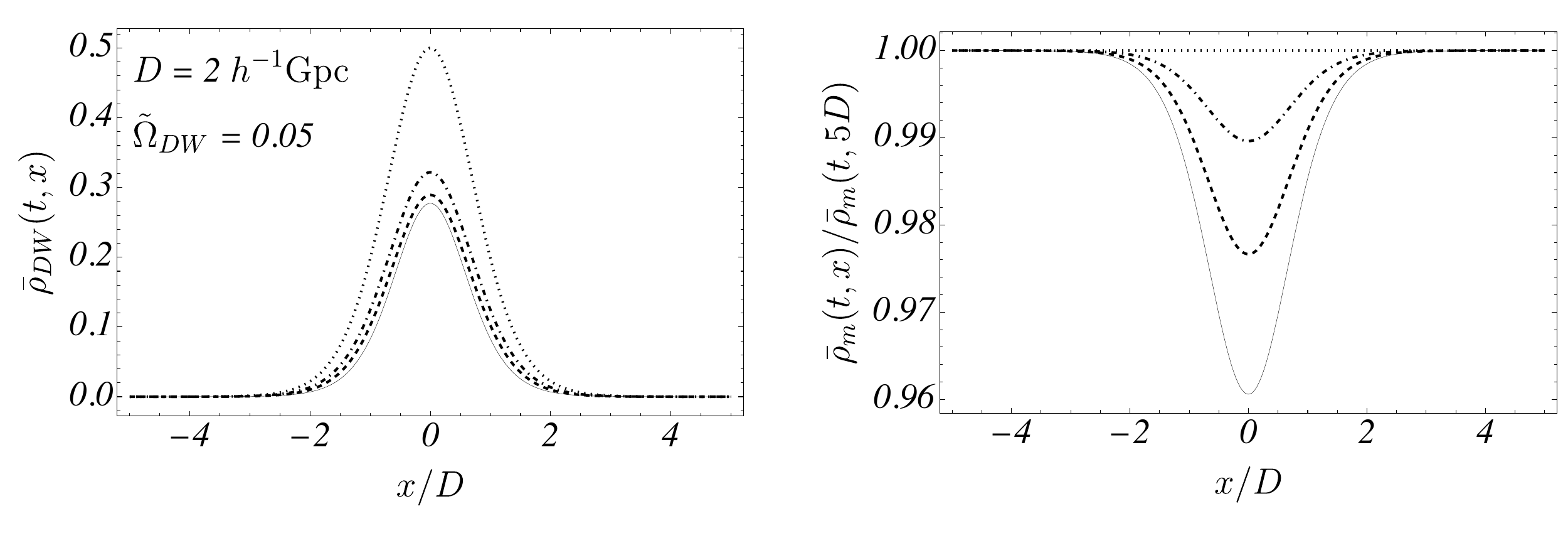}
        \end{minipage} \\
        \begin{minipage}{0.95\hsize}
            \centering
            \includegraphics[width=0.95\linewidth]{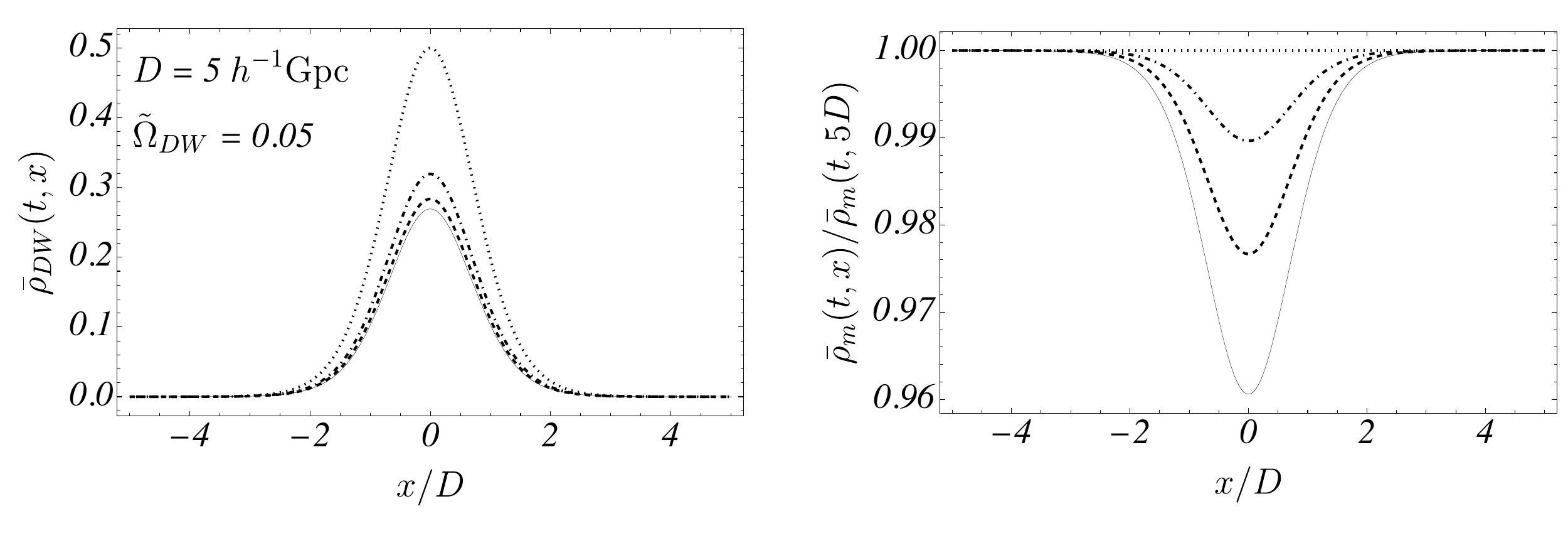}
        \end{minipage}
    \end{tabular}
    \caption{Energy densities of the DW (left) and matter (right)
    for $(D,
    \tilde{\Omega}_{\DW})=(2h^{-1}\,\si{Gpc}, 0.05)$ (top) and
    $(D,
    \tilde{\Omega}_{\DW})=(5h^{-1}\,\si{Gpc}, 0.05)$ (bottom). 
    The matter density is normalised by the value at $x=5D$, far away from the wall.
    The line styles are the same as figure~\ref{fig:phi}.
    As the energy density of the \ac{DW} becomes more pronounced, the matter energy density gets diluted around the centre.
    }
    \label{density}
\end{figure}

The energy densities of the \ac{DW} and matter are shown in figure~\ref{density}.  
It shows that as the energy density of \ac{DW} becomes more pronounced, the matter energy density is diluted around the centre.\footnote{It follows from eq.~\eqref{conserve1} that the matter velocity $v$ remains zero, starting from the vanishing velocity. Therefore, the \ac{DW} does not \emph{push} the matter \emph{out} but \emph{dilutes} it more via the enhanced Hubble parameter.}

\section{Geodesic equations}\label{Sec_geodesic}

In this work, we consider observational signatures of the \ac{DW} quintessence due to the bending of light paths via its energy density.
To investigate the light paths, 
we solve the geodesic equation~\cite{Alnes:2006pf,Enqvist:2006cg},
\begin{align}
    \dv[2]{x^\mu}{\lambda}+\Gamma^\mu_{\alpha\nu}\dv{x^\alpha}{\lambda}\dv{x^\nu}{\lambda}=0,
\end{align}
where $\lambda$ is the affine parameter, with its origin chosen such that $\lambda=0$ when the photon reaches the observer at time $t=t_0$, and 
\begin{align}
    \left.\dv{t}{\lambda}\right|_{t=t_0}=-1.
\end{align}
The negative sign reflects the fact that $\lambda$ increases as we trace the photon's path backwards in time from $t=t_0$.
\begin{figure}
\centering
\includegraphics[width=0.8\linewidth]{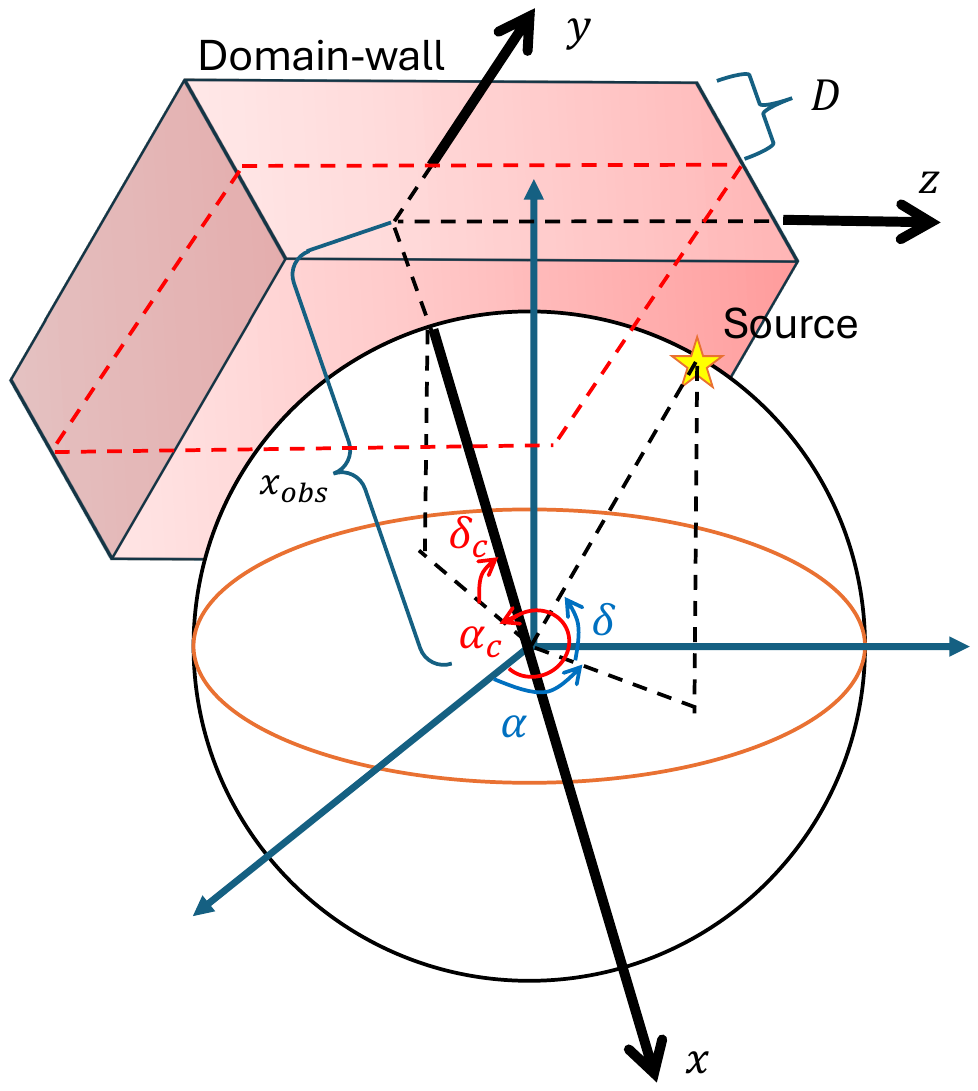}
\caption{Schematic illustration of the position of DW and light source on the equatorial coordinate system.
The observer is at the centre of the sphere.}
\label{wall_image}
\end{figure}
Thanks to the translational and rotational symmetry of the
spacetime in the $yz$ plane, we may, without loss of generality, place the observer on the $x$-axis and the source in the $xy$ plane.
Accordingly, we restrict our analysis to the two-dimensional $xy$ plane and define the observer's position as $(x,y) = (x_{\rm obs}, 0)$.

The geodesic equations for $\mu=t$, $x$, and $y$ read
\begin{align}
   (\mu=t)\quad &\dv[2]{t}{\lambda}+A\dot{A}\left(\dv{x}{\lambda}\right)^2+B\dot{B}\left(\dv{y}{\lambda}\right)^2=0,\\
   (\mu=x)\quad &\dv[2]{x}{\lambda}+\frac{A'}{A}\left(\dv{x}{\lambda}\right)^2-\frac{BB'}{A^2}\left(\dv{y}{\lambda}\right)^2+2\frac{\dot{A}}{A}\left(\dv{t}{\lambda}\dv{x}{\lambda}\right)=0,\\
   (\mu=y)\quad &\dv[2]{y}{\lambda}+2\frac{\dot{B}}{B}\left(\dv{t}{\lambda}\dv{y}{\lambda}\right)+2\frac{B'}{B}\left(\dv{x}{\lambda}\dv{y}{\lambda}\right)=0.
\end{align}
The first equation leads to the integral of motion,
\begin{align}
\label{4vel}
    -\left(\dv{t}{\lambda}\right)^2+A^2\left(\dv{x}{\lambda}\right)^2+B^2\left(\dv{y}{\lambda}\right)^2=0,
\end{align}
which can also be seen as the photon's 4-velocity identity, $u^\mu u_\mu=0$.
Defining $p_x \equiv \dv*{x}{\lambda}$ and 
$p_y \equiv \dv*{y}{\lambda}$, we can rewrite the system as the following set of five first-order differential equations:
\begin{align}
    \label{dxdl}
    \dv{x}{\lambda}&=p_x,\quad \dv{y}{\lambda}=p_y,\\
    \label{dtdl}
    \dv{t}{\lambda}&=-\sqrt{A^2p_x^2+B^2p_y^2},\\
    \dv{p_x}{\lambda}&=-\frac{A'}{A}p_x^2+\frac{BB'}{A^2}p_y^2+2\frac{\dot{A}}{A}p_x\sqrt{A^2p_x^2+B^2p_y^2},\\
    \label{dpydl}
    \dv{p_y}{\lambda}&=-2\frac{B'}{B}p_xp_y+2\frac{\dot{B}}{B}p_y\sqrt{A^2p_x^2+B^2p_y^2}.
\end{align}

The unit vector $v^i$ in the $x$-direction and the unit tangent vector $\bar{u}^i$ to the photon's geodesic are given by
\begin{align}
    v^i = \frac{1}{A}(1, 0),\quad
    \bar{u}^i = \left|\dv{\lambda}{t}\right|(p_x, p_y).
\end{align}
The angle $\xi$ between the photon's trajectory and the $x$-axis, as seen by the observer, is given by the inner product of $\bar{u}^i$ and $v^i$
\begin{align}
    \cos\xi = g_{ij}\bar{u}^i v^j|_{t=t_0} 
    &= A(x_{\rm obs}, t_0)p_x(x_{\rm obs}, t_0)
    \label{cdot}.
\end{align}
Thus, we have
\begin{align}
\label{px0}
    p_x(x_{\rm obs}, t_0) = \frac{\cos\xi}{A(x_{\rm obs}, t_0)}.
\end{align}
Substituting it into eq.~\eqref{dtdl}, one finds the initial value of $p_y$ as
\begin{align}
    p_y(x_{\rm obs}, t_0) = \frac{\sin\xi}{B(x_{\rm obs}, t_0)}.
\end{align}
It should also be noted that
eq.~(\ref{dpydl}) can be rewritten as
\begin{align}
     \dv{p_y}{\lambda} 
     &= -2p_y \dv{\lambda}\ln B(t(\lambda),x(\lambda)),
\end{align}
implying that
$B^2p_y$ is constant along the geodesic.

Next, we define the redshift 
$\mathsf{z}$ of a photon received by the observer 
(take care not to confuse it with the third spatial coordinate $z$).
It is defined as the ratio of the time intervals of photon pulses between their emissions and receptions.
Assume two photons are emitted from the source with a time interval $\Delta t$. Let the time component of the first photon be $t_1 = t(\lambda)$, and that of the second be $t_2 = t(\lambda) + \Delta t(\lambda)$. Since both photons satisfy the 4-velocity identity~(\ref{4vel}), we have for the first photon,
\begin{align}
    \label{photon1}
    -\left(\dv{t}{\lambda}\right)^2+A^2(x,t)\left(\dv{x}{\lambda}\right)^2+B^2(x,t)\left(\dv{y}{\lambda}\right)^2=0.
\end{align}
and for the second photon,
\begin{align}
    \label{photon2}
    -\left(\dv{(t+\Delta t)}{\lambda}\right)^2+A^2(x,t+\Delta t)\left(\dv{x}{\lambda}\right)^2+B^2(x,t+\Delta t)\left(\dv{y}{\lambda}\right)^2=0.
\end{align}
Expanding (\ref{photon2}) to first order in $\Delta t$ and using (\ref{photon1}), we obtain
\begin{align}
\label{dt}
    \dv{t}{\lambda}\dv{\Delta t}{\lambda} = \Delta t(\lambda)\left[A\dot{A}\left(\dv{x}{\lambda}\right)^2 + B\dot{B}\left(\dv{y}{\lambda}\right)^2\right].
\end{align}
The redshift is then defined as
\begin{align}
\label{red}
    1 + \mathsf{z}(\lambda_\ue) \equiv \frac{\Delta t(\lambda_\ur)}{\Delta t(\lambda_\ue)},
\end{align}
where the subscripts `e' and `r' of the affine parameter denote emission and reception, respectively. Differentiating eq.~(\ref{red}) with respect to $\lambda_\ue$ using eq.~(\ref{dt}) and dropping the subscript `e' for brevity, we have
\begin{align}
    \label{dzdl}
    \dv{\mathsf{z}}{\lambda} = \frac{A\dot{A}p_x^2 + B\dot{B}p_y^2}{\sqrt{A^2p_x^2 + B^2p_y^2}}(1+\mathsf{z}).
\end{align}
This equation describes how the redshift changes along a small increment of the affine parameter $\dd{\lambda}$. To compute the redshift of a photon reaching the observer at present, one can integrate this equation along the past light cone.

In summary, we solve
first-order system eqs.~(\ref{dxdl})--(\ref{dpydl}) and eq.~(\ref{dzdl})
with the initial conditions,
\begin{align}
    t(\lambda=0)=t_0,\quad
    x(\lambda=0)=x_{\rm obs},\quad
    y(\lambda=0)=0,\\
    p_x(\lambda=0)=\frac{\cos\xi}{A(x_{\rm obs}, t_0)},\quad
    p_y(\lambda=0)=\frac{\sin\xi}{B(x_{\rm obs}, t_0)}.
\end{align}
By solving this system, we can determine the photon's geodesic,
\begin{align}
    t(\lambda, \xi, x_{\rm obs}), \quad x(\lambda, \xi, x_{\rm obs}), \quad
    y(\lambda, \xi, x_{\rm obs}), \quad \mathsf{z}(\lambda, \xi, x_{\rm obs}).
\end{align}
These solutions also allow us to treat $t$, $x$, and $y$
as implicit functions of the redshift $\mathsf{z}$, which are useful for later analyses.

\begin{figure}
  \centering
  \begin{minipage}{0.4\columnwidth}
     \centering
     \includegraphics[width=\columnwidth]{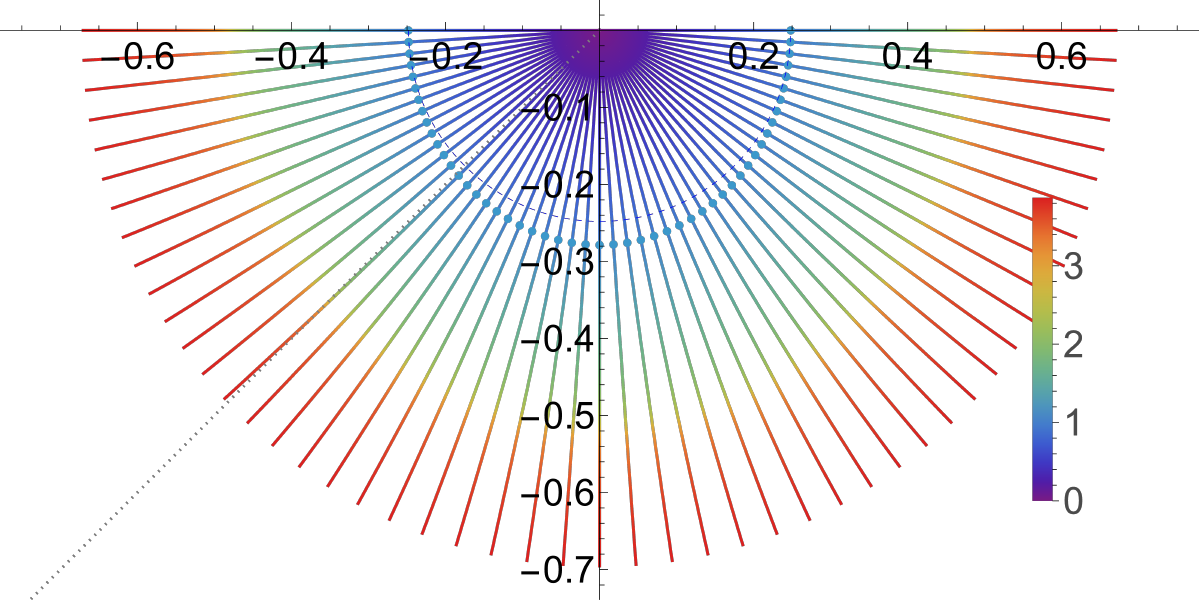}
  \end{minipage}
 \begin{minipage}{0.4\columnwidth}
     \centering
     \includegraphics[width=\columnwidth]{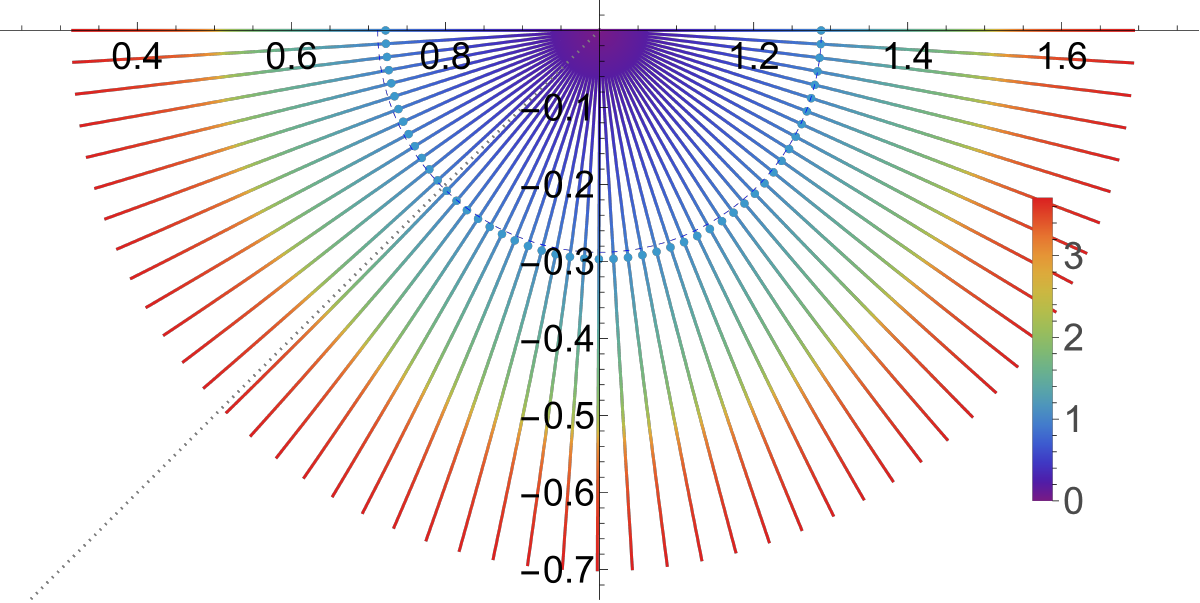}
  \end{minipage}
  \begin{minipage}{0.4\columnwidth}
     \centering
     \includegraphics[width=\columnwidth]{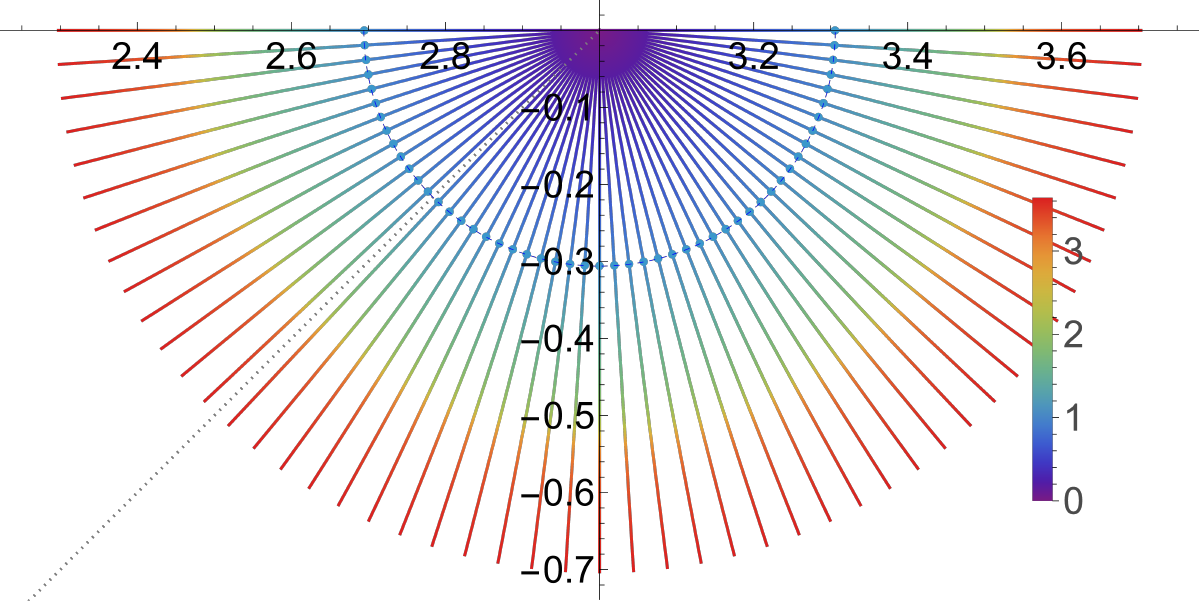}
  \end{minipage}
\caption{Geodesics
in the case $(D,\,\tilde{\Omega}_{\DW})=(2h^{-1}\,\si{Gpc},\, 0.2)$ (note that the parameter choice is not physically realistic and is adopted only to exaggerate the effect for illustrative purposes) for $\bar{x}_{\rm obs}=0$ (upper left), $\bar{x}_{\rm obs}=1$ (upper right), and
$\bar{x}_{\rm obs}=3$ (bottom). The horizontal axis shows the normalised $x$-coordinate, and the vertical axis shows the $y$-coordinate. The black dotted line is a straight auxiliary line. The blue dots are the positions where the redshift $\mathsf{z}=1.0$. The blue dashed line is the semicircle as an auxiliary line.}
\label{geodesic}
\end{figure}

Figure~\ref{geodesic} shows the geodesics of light rays reaching the observer, colour-coded by redshift,
in the case of $(D,\,\tilde{\Omega}_{\DW})=(2h^{-1}\,\si{Gpc},\, 0.2)$, with varying values of $x_{\rm obs}$. When the observer is located near the DW, the bending of the geodesics is obvious. To make the curvature more visually clear, each panel also includes a dashed straight line for comparison.
In contrast, when the observer is farther away, the geodesics are nearly straight, indicating that the light travels along a nearly straight path. This behaviour can be attributed to the significant spacetime distortion near the DW, while at large distances the spacetime asymptotically approaches that of a $\Lambda$CDM universe. Focusing on the geodesic for $\bar{x}_{\rm obs} = 1$, we observe that the surfaces of a constant redshift are stretched in the direction away from the DW and compressed toward the near side. As seen in figure~\ref{AB}, this is because the expansion rate of space is greater near the central wall, so photons arriving from that direction cover shorter distances for the same redshift. Furthermore, examining the distortion of the constant-redshift surfaces, we find that near the DW, the redshift contours become elliptically elongated in the $y$-direction compared to the $x$-direction, while at large distances they tend to become semicircular. The blue dots are the positions where the redshift $\mathsf{z}=1$, and the blue dashed line is a semicircle for reference. This illustrates that the spacetime is anisotropic near the DW and becomes isotropic farther away.

\section{Constraints from the CMB}\label{constraintCMB}

The bending of photon paths due to the \ac{DW} yields anisotropies in the  \ac{CMB} in addition to the primordial perturbation.
In this section, we evaluate the constraints on the model parameters from these anisotropies.
To obtain a conservative bound, we neglect the primordial perturbation and consider anisotropies induced purely by the \ac{DW}.
Let us suppose that the initial time $t_\ui$ of our simulation is taken to be sufficiently early and the \ac{CMB} anisotropies are mainly produced after this time.
Although we do not explicitly compute the effects from the last scattering surface to the initial time,
our estimate of the anisotropy is expected to be conservative in the following sense. During this early propagation epoch, the \ac{DW} energy density is much smaller than its late-time value, and hence the \ac{DW}-induced distortion of photon geodesics and redshifts should be subdominant. We therefore expect the dominant contribution to the anisotropy to be generated after the initial time, which is the regime captured by our numerical calculation.
The accumulated anisotropies are hence encoded in the direction-dependent redshift at $t_\ui$. The initial affine parameter $\lambda_\ui$ is first defined as
$t_\ui = t(\lambda_\ui, \mathbf{s}, \xi, x_{\mathrm{obs}})$, and then the redshift is calculated as
$\mathsf{z}_\ui(\mathbf{s}, \xi, x_{\mathrm{obs}}) = \mathsf{z}(\lambda_\ui, \mathbf{s}, \xi, x_{\mathrm{obs}})$,
where $\mathbf{s} = (D, \tilde{\Omega}_\DW)$.
The observable anisotropic temperature is given by
\begin{align}
    T(\mathbf{s}, \xi, x_{\mathrm{obs}}) =
    \frac{T_\ui}{1 + \mathsf{z}_\ui(\mathbf{s}, \xi, x_{\mathrm{obs}})},
\end{align}
where $T_\ui$ is the isotropic temperature at initial time $t_\ui$.

Taking the angular average of the observable temperature, we obtain the present isotropic temperature as
\begin{align}
  T_0(\mathbf{s}, x_{\mathrm{obs}})
  = \frac{1}{4\pi}\int \dd{\Omega} T(\mathbf{s}, \xi, x_{\mathrm{obs}})
  = \frac{T_\ui}{2}\int_{0}^{\pi}
  \frac{\sin\xi}{1 + \mathsf{z}_\ui(\mathbf{s}, \xi, x_{\mathrm{obs}})}\dd{\xi}.
\end{align}
Therefore, the temperature perturbation caused by the DW is
\begin{align}
    \frac{\Delta T}{T_0}(\mathbf{s}, \xi, x_{\mathrm{obs}})
    \equiv
    \frac{T(\mathbf{s}, \xi, x_{\mathrm{obs}}) - T_0(\mathbf{s}, x_{\mathrm{obs}})}
    {T_0(\mathbf{s}, x_{\mathrm{obs}})}
    =
    \frac{\bar{\mathsf{z}}(\mathbf{s}, x_{\mathrm{obs}}) - \mathsf{z}_\ui(\mathbf{s}, \xi, x_{\mathrm{obs}})}
    {1 + \mathsf{z}_\ui(\mathbf{s}, \xi, x_{\mathrm{obs}})},
    \label{temper}
\end{align}
where
\begin{align}
    \bar{\mathsf{z}}(\mathbf{s}, x_{\mathrm{obs}})
    \equiv
    2\left(
    \int_0^\pi \dd{\xi} \frac{\sin\xi}{1 + \mathsf{z}_\ui(\mathbf{s}, \xi, x_{\mathrm{obs}})}
    \right)^{-1} - 1 \, .
\end{align}
The temperature perturbation can be expanded to spherical harmonics as
\begin{align}
    \frac{\Delta T}{T_0}(\mathbf{s}, \xi, x_{\mathrm{obs}})
    =
    \sum_{l,m}
    a_{lm}(\mathbf{s}, x_{\mathrm{obs}})
    Y_{lm}(\xi, \phi),
\end{align}
and the predicted multipole coefficients are given by
\begin{align}
    a_{lm}(D, \tilde{\Omega}_\DW, x_{\mathrm{obs}})
    =
    \int_0^{2\pi} \dd{\phi}
    \int_0^\pi \dd{\xi}
    \frac{\Delta T}{T_0}(\mathbf{s}, \xi, x_{\mathrm{obs}})
    Y_{lm}^*(\xi, \phi)
    \sin\xi,
\end{align}
where the asterisk denotes complex conjugation.
Due to the axial symmetry of the DW, the temperature perturbation also has axial symmetry, so all coefficients vanish except for $m = 0$.

\subsection{Dominant cosmological constant and subdominant DW case}\label{CMB}

\begin{figure}
    \centering
    \begin{tabular}{c}
    \begin{minipage}{0.45\columnwidth}
        \centering
        \includegraphics[width=0.95\linewidth]{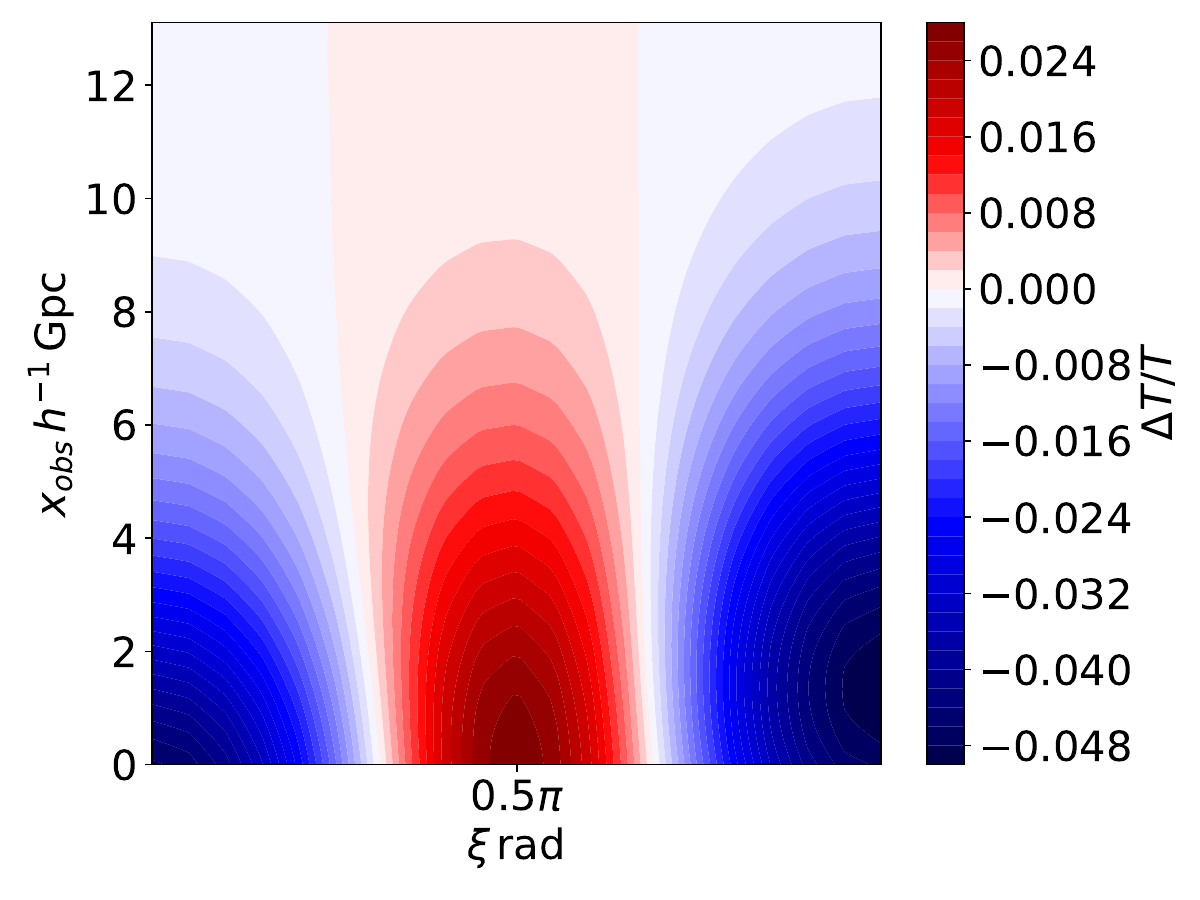}
    \end{minipage}
    \begin{minipage}{0.45\columnwidth}
        \centering
        \includegraphics[width=0.95\linewidth]{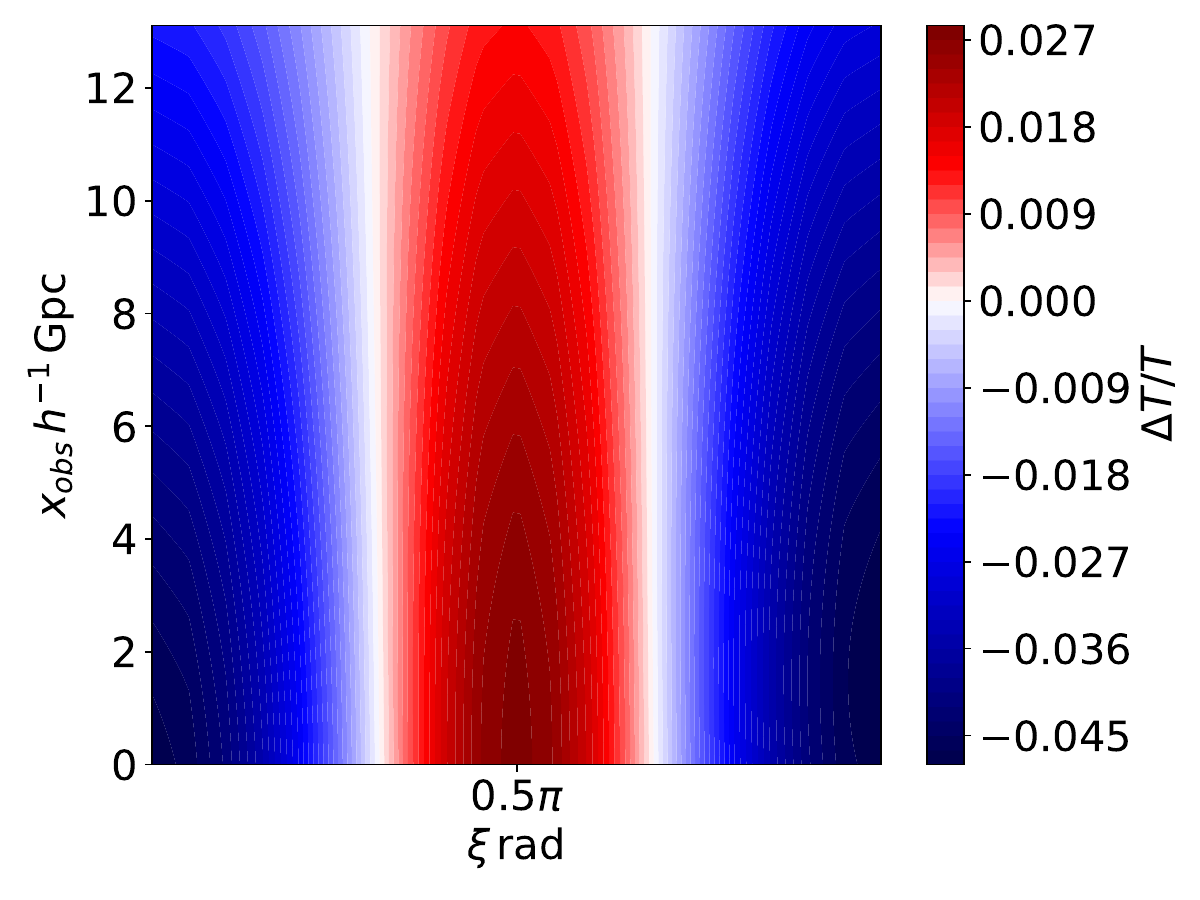}
    \end{minipage}
    \end{tabular}
\caption{Temperature perturbation induced by the DW for $\tilde{\Omega}_{\DW} = 0.2$ with $D = 5h^{-1}\,\si{Gpc}$ (left) and $D = 15h^{-1}\,\si{Gpc}$ (right).
The temperature perturbation remains significant even at large $x_{\mathrm{obs}}$ for thicker DW.}
\label{tf}
\end{figure}

\begin{figure}
  \centering
  \begin{minipage}{0.4\columnwidth}
     \centering
     \includegraphics[width=\columnwidth]{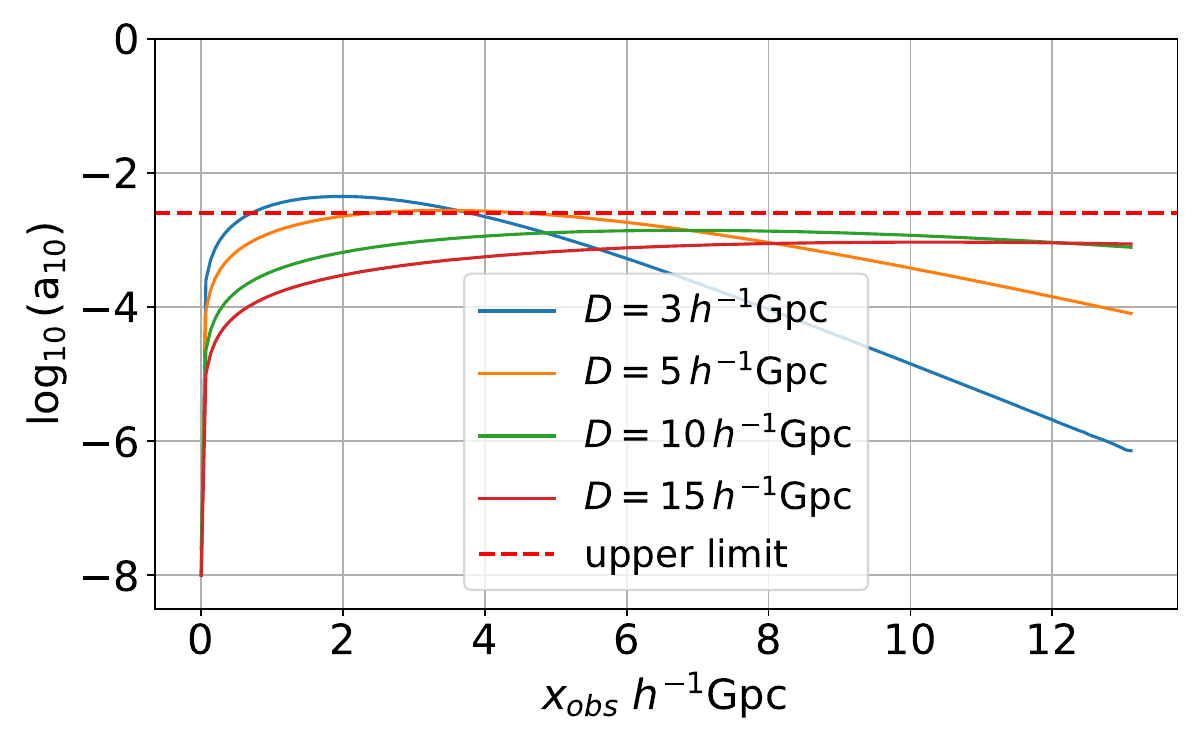}
  \end{minipage}
  \begin{minipage}{0.4\columnwidth}
     \centering
     \includegraphics[width=\columnwidth]{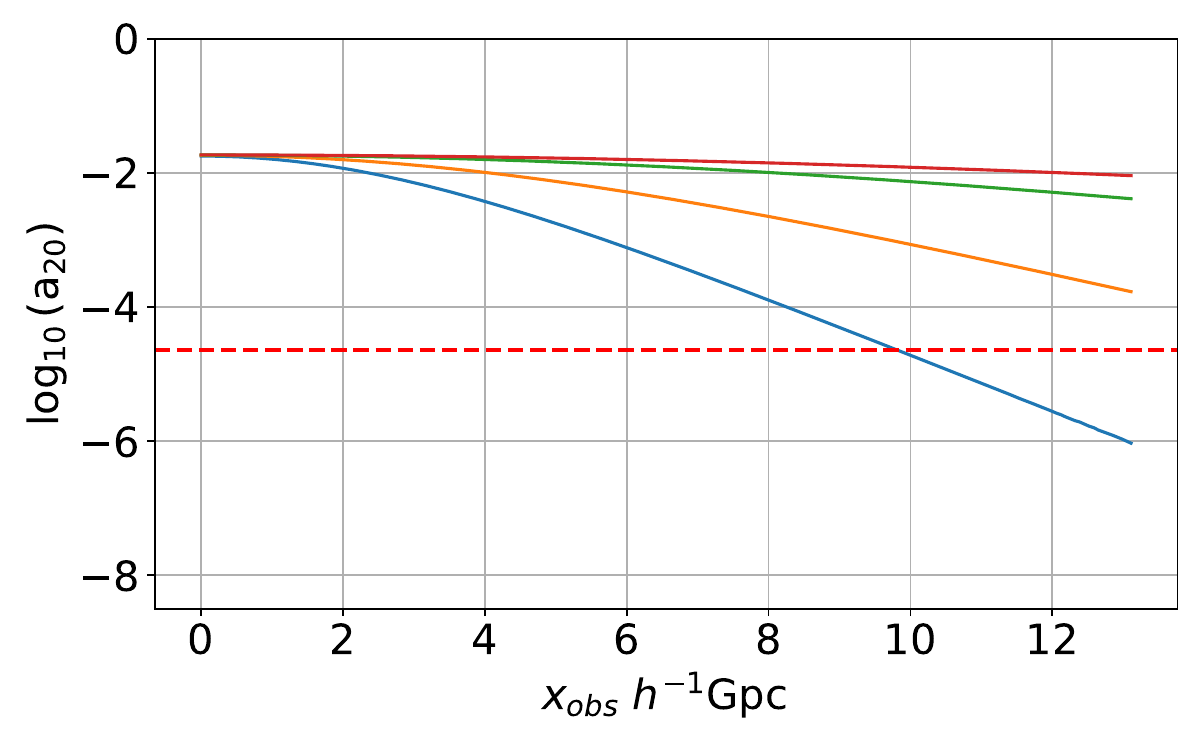}
  \end{minipage}
  \begin{minipage}{0.4\columnwidth}
     \centering
     \includegraphics[width=\columnwidth]{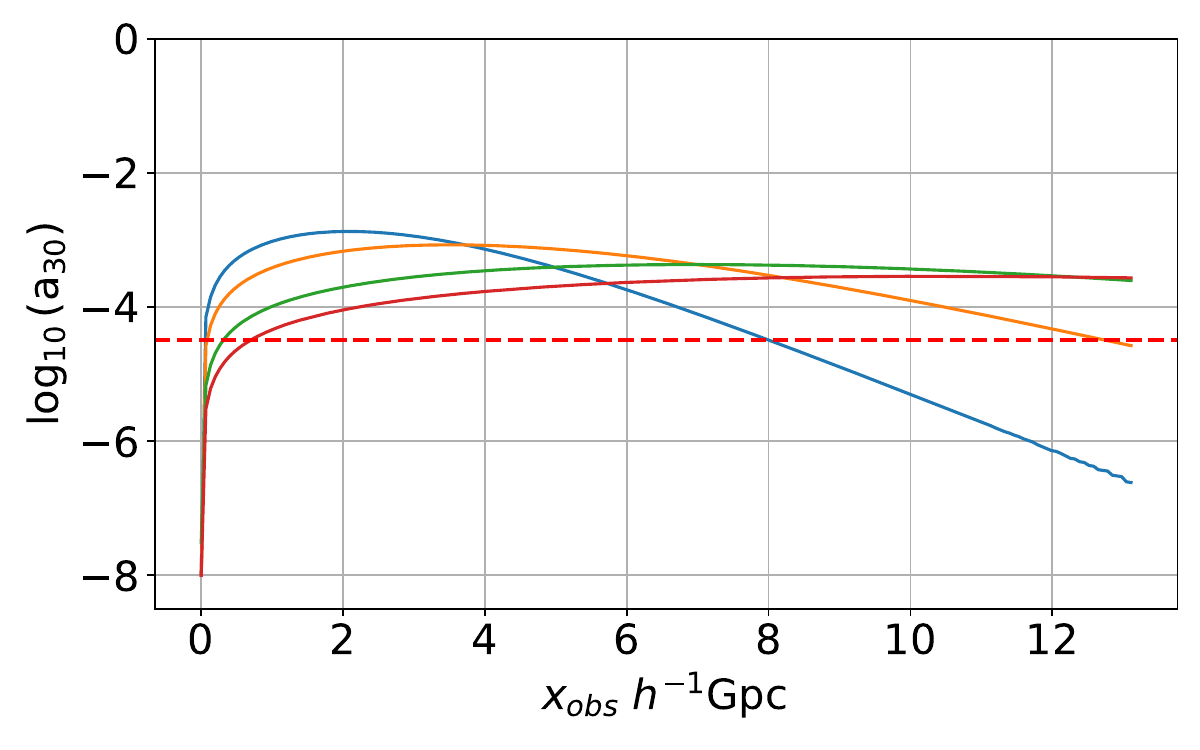}
  \end{minipage}
  \caption{
Predicted multipole moments (log scale) for $\tilde{\Omega}_{\DW} = 0.05$.
Upper left: dipole amplitude $a_{10}$. Upper right: quadrupole amplitude $a_{20}$. Bottom: octopole amplitude $a_{30}$. The red dashed line shows the upper bound of the Planck 2018 observation results.
At the DW centre ($x_{\rm obs}=0$), the temperature distribution is symmetric with respect to $\xi<\pi/2$ and $\xi>\pi/2$, causing odd multipoles to vanish, while the contrast between directions perpendicular and parallel to the DW is maximal, yielding the largest quadrupole.
Far from the DW (large $x_{\rm obs}$), the temperature field becomes nearly isotropic, and all multipole components approach zero.
}
\label{dipquad}
\end{figure}

Figure~\ref{tf} shows the temperature perturbation for $\tilde{\Omega}_{\DW} = 0.2$.
Temperature perturbations remain detectable even for large $x_{\mathrm{obs}}$ for a thicker DW (i.e., larger $D$).
Figure~\ref{dipquad} shows the predicted dipole, quadrupole, and octopole moments for $\tilde{\Omega}_{\DW} = 0.2$.
At the centre of the domain wall ($x_{\rm obs}=0$), the temperature distribution is symmetric with respect to $\xi<\pi/2$ and $\xi>\pi/2$ (see figure~\ref{tf}), so odd multipoles, such as the dipole and octopole, vanish. 
In contrast, the temperature difference between directions perpendicular and parallel to the DW is maximal, which makes the quadrupole nonzero and dominant. 
Far from the DW (large $x_{\rm obs}$), the temperature field becomes nearly isotropic, and all multipole components approach zero. 
When the observer is slightly displaced from the centre, a left-right asymmetry arises, generating odd multipoles such as the dipole and octopole.

\begin{figure}
    \centering
    \begin{tabular}{c}
    \begin{minipage}{0.45\columnwidth}
        \centering
        \includegraphics[width=0.95\linewidth]{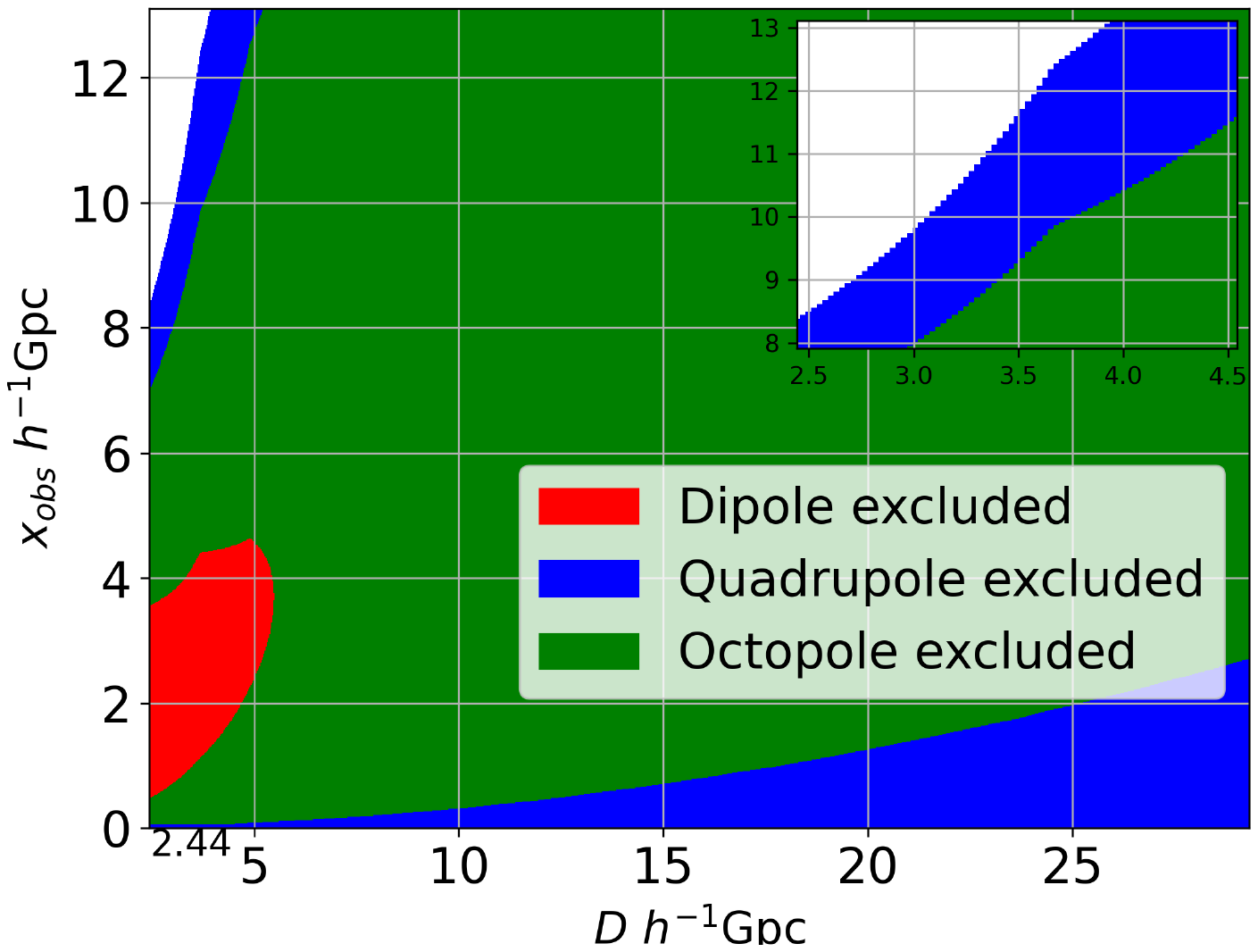}
    \end{minipage}
    \begin{minipage}{0.45\columnwidth}
        \centering
        \includegraphics[width=0.95\linewidth]{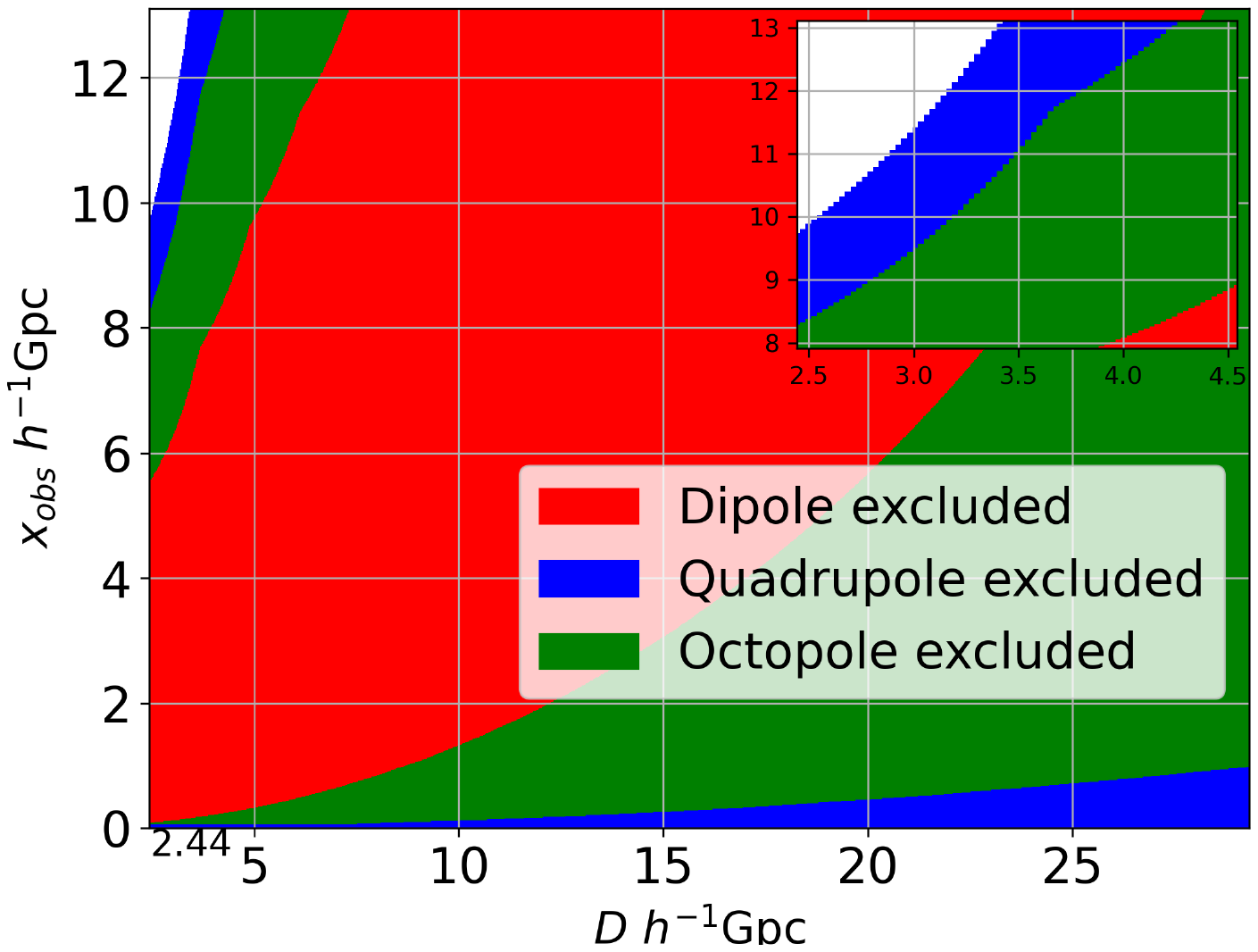}
    \end{minipage}
    \end{tabular}
\caption{Excluded parameter regions from the CMB dipole (red), quadrupole (blue) and octopole (green) constraints.
The white region indicates the parameter space consistent with all observational multipole bounds.
The left panel is for $\tilde{\Omega}_{\DW}=0.05$ and the right panel is for $\tilde{\Omega}_{\DW}=0.2$. These results favour the regime with small $D$ and large
$x_{\mathrm{obs}}$ where the DW-induced anisotropies are suppressed.
}
\label{multipole}
\end{figure}

We use the Planck 2018 results to constrain the model parameters $(D, x_{\mathrm{obs}})$.
The measured CMB dipole amplitude is~\cite{Planck:2018nkj}
\begin{align}
    a_{10}^{\mathrm{obs}} = \sqrt{\frac{4\pi}{3}}(1.23357 \pm 0.00036)\times10^{-3}.
\end{align}
The predicted dipole must be smaller than this observational upper bound, so we adopt this upper bound as a constraint.
To formulate the quadrupole and octopole constraints, we define the angular power spectrum as
\begin{align}
  \langle a_{lm} a_{l'm'}^* \rangle
  T_0^2=
  \delta_{ll'}\delta_{mm'} C_l.
  \end{align}
In our model only $m=0$ survives, so
\begin{align}
    C_l = \frac{|a_{l0}|^2}{2l+1} T_0^{\,2}.
\end{align}
The measured quadrupole and octopole powers of the CMB, taken from the $1\sigma$ upper bound of the observational error bars, are~\cite{Planck:2018vyg}
\begin{align}
    C_{2,\mathrm{obs}}
    \lesssim 795\,\si{\mu K^2}, \quad
    C_{3,\mathrm{obs}} 
    \lesssim 1125\,\si{\mu K^2}.
\end{align}
Thus, the observational upper bounds on $a_{20}$ and $a_{30}$ are
\begin{align}
    a_{20}^{\mathrm{obs}}
    =
    \sqrt{
    \frac{5\,C_{2,\mathrm{obs}}}
    {T_0^2}
    }
    \le 2.31\times10^{-5}, \quad 
    a_{30}^{\mathrm{obs}}
    =
    \sqrt{
    \frac{7\,C_{3,\mathrm{obs}}}
    {T_0^2}
    }
    \le 3.26\times10^{-5},
\end{align}
where $T_0=2.7255~\si{K}$.
The regions excluded by these multipole constraints are shown in figure~\ref{multipole}.
Only a parameter region with small $D$ and large $x_{\mathrm{obs}}$ (i.e., the homogeneous and isotropic limit) remains allowed. 

The parameter region allowed by figure~\ref{multipole} satisfies\footnote{Evaluating eq.~(\ref{condition}) at the initial time, we obtain $\rho_{\DW}\lesssim1\times10^{-4}\rho_{\rm{crit}}\approx(0.3h^{1/2}\,\si{meV})^4$.}
\begin{align}
\label{condition}
\rho_{\rm DW}(t_0,x_\obs) \lesssim (1\text{--}2)\times
10^{-5}
\rho_{\mathrm{crit}}
\qquad (\text{for } \tilde{\Omega}_{\rm DW} \lesssim 0.15),
\end{align}
where $\rho_{\mathrm{crit}}$ is the present critical density, and
$x_{\mathrm{obs}}$ is located far away from the \ac{DW}, in the tail region. 
This implies that the \ac{DW} is subdominant in the present universe.

\subsection{DW-dominant case}\label{pure}

\begin{figure}
\centering
\includegraphics[width=0.8\linewidth]{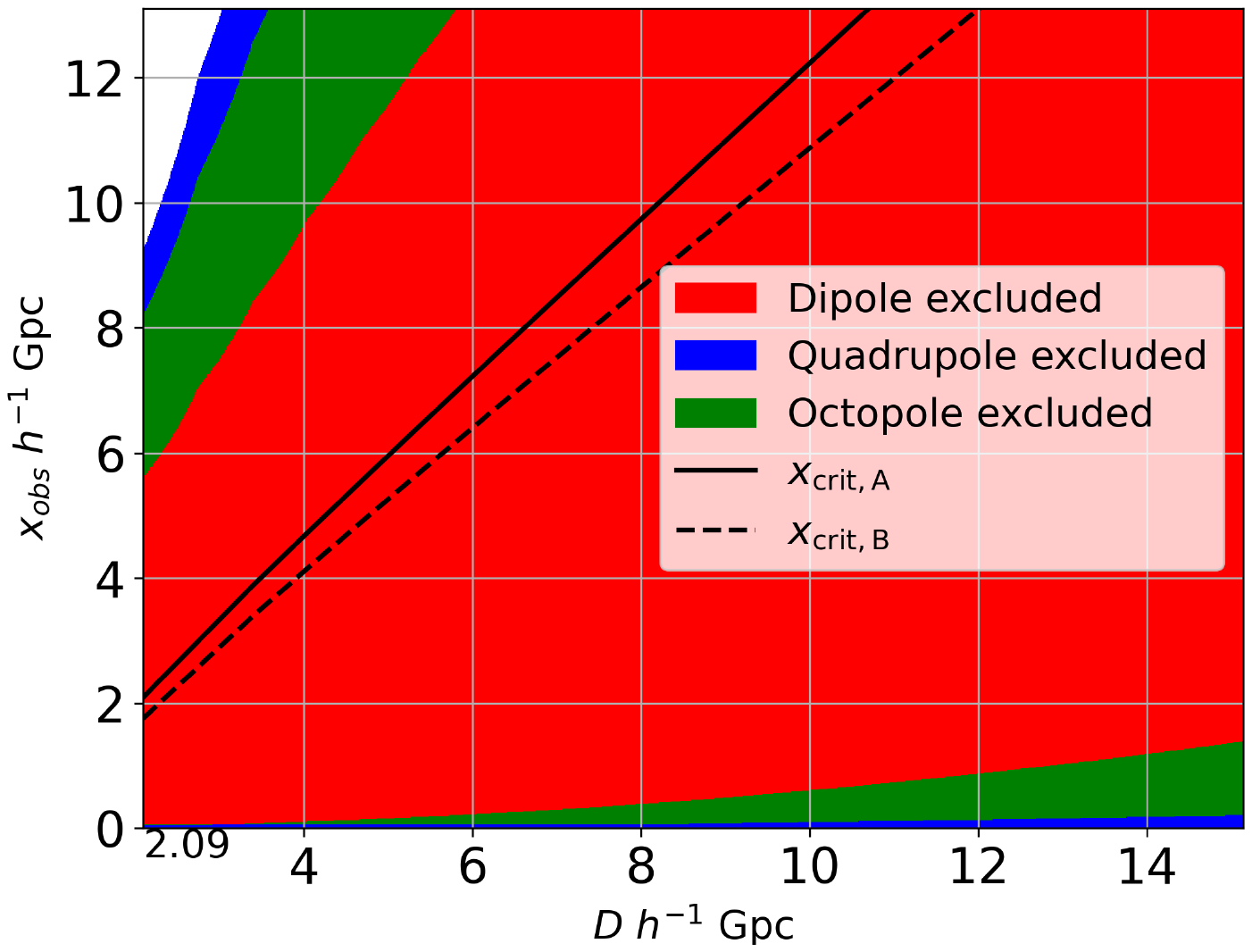}
\caption{
Excluded parameter regions from the CMB constraint for the case 
$\tilde{\Omega}_{\DW}=0.7$ and $\Omega_{\Lambda}=0$. 
The colour scheme is the same as in figure~\ref{multipole}. 
The region below the black solid line corresponds to parameters that realise accelerated expansion for the scale factor $A$, while the black dashed line corresponds to that for the scale factor $B$. 
This result shows that pure domain-wall dark energy cannot simultaneously satisfy the CMB anisotropy constraint and the requirement of accelerated expansion.
}
\label{pureDW}
\end{figure}

We also consider the case in which dark energy is purely composed of a DW.
Such a scenario could be motivated by applying the topological inflation mechanism to dark energy. 
We define $x_{\mathrm{crit},A}$ and $x_{\mathrm{crit},B}$ as
\begin{align}
    \eval{\frac{\partial^2 A(t
    ,x_{\mathrm{crit},A})}{\partial t^2}}_{t=t_0} = 0
    \qc
    \eval{\frac{\partial^2 B(t,x_{\mathrm{crit},B})}{\partial t^2}}_{t=t_0} = 0 .
\end{align}
Thus, regions with 
$x<x_{\mathrm{crit},A}$ and 
$x < x_{\mathrm{crit},B}$ undergo an accelerated expansion at the present.
Figure~\ref{pureDW} shows the case with $\tilde{\Omega}_{\DW}=0.7$ and $\Omega_\Lambda=0$. 
We find that there are no regions that simultaneously satisfy the CMB anisotropy constraint and the condition for accelerated expansion at present. 
This result indicates a no-go for the pure planar DW dark energy model.

\section{Constraints from the SNe Ia Data}\label{MCMC}

In this section, we evaluate the parameter constraints using the Type Ia \acp{SN} data such as Pantheon+~SH0ES~\cite{Riess:2021jrx,Brout:2022vxf} and \ac{DESY5}~\cite{DES:2025sig}.

Due to the presence of the DW, spacetime is curved, which affects the propagation of light. In such a spacetime,
the angular diameter distance $d_A$ satisfies~\cite{ELLIS1985315,Alnes:2006uk}
\begin{align}
    d_A^4 \sin^2\xi = \tilde{g}_{\xi\xi}\tilde{g}_{\phi\phi} - (\tilde{g}_{\xi\phi})^2,
    \label{dA}
\end{align}
where $\tilde{g}_{\mu\nu}$ is the metric in the observer frame, a coordinate system defined along past-directed null geodesics and $\mu,\nu=0,1,2,3$. Its coordinates are $(\mathsf{t_0}, t - \mathsf{t_0}, \xi, \phi)$, with $\xi$ and $\phi$ representing the zenith and azimuthal angles of the photon at the observation time $\mathsf{t_0}$.
In the following analysis, we consider observations at present, i.e., we set  $\mathsf{t_0} = t_0$.

Transforming from the observer frame to the metric of eq.~(\ref{metric}), we have
\begin{align}
    t = t,\quad x^i = x^i(t_0, t - t_0, \xi, \phi),
\end{align}
where $i=1,2,3$.
The observer-frame metric is related to the original metric $g_{\mu\nu}$ via
\begin{align}
    \tilde{g}_{\mu\nu} = g_{\sigma\tau} \frac{\partial x^\sigma}{\partial \tilde{x}^\mu} \frac{\partial x^\tau}{\partial \tilde{x}^\nu}.
\end{align}
To compute $\tilde{g}_{\phi\phi}$, we use $\partial x / \partial \phi = 0$, $\partial y / \partial \phi = -z$, and $\partial z / \partial \phi = y$ (note that $x$ is the zenith direction in our definition). For $\tilde{g}_{\xi\phi}$, we assume $(\partial y / \partial \xi, \partial z / \partial \xi) \propto (y, z)$. Furthermore, assuming that a SN Ia lies in the $xy$-plane ($z=0$) implies that $\xi$ coincides with the angle defined in eq.~(\ref{cdot}) and $\phi = 0$. Therefore, the observer frame metric simplifies to
\begin{align}
    \tilde{g}_{\xi\xi} &= A^2\left(\frac{\partial x}{\partial \xi}\right)^2 + B^2\left(\frac{\partial y}{\partial \xi}\right)^2, \\
    \tilde{g}_{\phi\phi} &= B^2 y^2, \quad \tilde{g}_{\xi\phi} = 0.
\end{align}
Substituting them into eq.~(\ref{dA}), we obtain a simplified expression for the angular diameter distance:
\begin{align}
\label{dAsimple}
    d_A^4 \sin^2\xi = \left[A^2\left(\frac{\partial x}{\partial \xi}\right)^2 + B^2\left(\frac{\partial y}{\partial \xi}\right)^2\right] B^2 y^2.
\end{align}
In the
Friedmann--Lema\^itre--Robertson--Walker limit ($A \sim B \sim a(t)$), we recover $d_A = a(t)r$ with $x = r\cos\xi$, $y = r\sin\xi\cos\phi$, $z = r\sin\xi\sin\phi$. The Etherington's reciprocity theorem relates the luminosity distance $d_L$ to the angular diameter distance: $d_L = (1 + \mathsf{z})^2 d_A$.
Thus, eq.~(\ref{dAsimple}) gives
\bme{
    d_L^4(\mathbf{d}) = (1 + \mathsf{z})^8 \frac{B^2(x(\mathbf{d}), t(\mathbf{d}))\, y^2(\mathbf{d})}{\sin^2\xi} \\
    \times\left[
    A^2(x(\mathbf{d}), t(\mathbf{d})) \left( \frac{\partial x(\mathbf{d})}{\partial \xi} \right)^2 +
    B^2(x(\mathbf{d}), t(\mathbf{d})) \left( \frac{\partial y(\mathbf{d})}{\partial \xi} \right)^2
    \right],
}
where $\mathbf{d} \equiv (D, \tilde{\Omega}_{\DW}, h, \mathsf{z}, \xi, x_{\rm obs})$. The derivatives with respect to $\xi$ are computed numerically along the geodesics.

The distance modulus $\mu$ of an \acp{SN} Ia is then uniquely determined by the luminosity distance as (see, e.g., Ref.~\cite{Grande:2011hm})
\begin{align}
    \mu(\mathbf{d})= 5\log_{10}\left(\frac{d_L(\mathbf{d})}{1h^{-1}\,\si{Gpc}}\right) - 5\log_{10}h + 40.
    \label{mu2}
\end{align}
To compare with SNe Ia observations, we relate the celestial coordinates of each SN Ia to the direction normal to the DW surface via $\xi$ (see figure~\ref{wall_image}). In the equatorial coordinate system, a celestial object is located at right ascension $\alpha$ and declination $\delta$, while the DW normal is at $(\alpha_\cc, \delta_\cc)$. The unit vector $v^i$ from the DW to the observer, and the unit tangent vector $\bar{u}^i$ of the SNe Ia geodesic at present are
\begin{align}
     v^i = -
     \begin{pmatrix}
     \sin(90^\circ - \delta_\cc)\cos\alpha_\cc \\
     \sin(90^\circ - \delta_\cc)\sin\alpha_\cc \\
     \cos(90^\circ - \delta_\cc)
     \end{pmatrix},\quad
     \bar{u}^i =
     \begin{pmatrix}
     \sin(90^\circ - \delta)\cos\alpha \\
     \sin(90^\circ - \delta)\sin\alpha \\
     \cos(90^\circ - \delta)
     \end{pmatrix}.
\end{align}
The angle $\xi$ between these vectors is
\begin{align}
    \cos\xi =  -\sin(90^\circ - \delta_\cc)\sin(90^\circ - \delta)\cos(\alpha + (360^\circ - \alpha_\cc))
      - \cos(90^\circ - \delta_\cc)\cos(90^\circ - \delta),
\end{align}
so that $\xi = \xi(\alpha, \delta, \alpha_\cc, \delta_\cc)$.
Finally, substituting this into eq.~(\ref{mu2}) expresses the distance modulus as a function of nine parameters $\mu = \mu(D, \tilde{\Omega}_{\DW}, h, \mathsf{z}, \alpha, \delta, \alpha_\cc, \delta_\cc, x_{\rm obs})$.
Using observed SNe Ia data $\mu(\mathsf{z}, \alpha, \delta)$, we can estimate the following six parameters $\{D, \tilde{\Omega}_{\DW}, h, \alpha_\cc, \delta_\cc, x_{\rm obs}\}$.

\begin{figure}
    \centering
    \begin{tabular}{c}
    \begin{minipage}{0.45\columnwidth}
        \centering
        \includegraphics[width=0.95\linewidth]{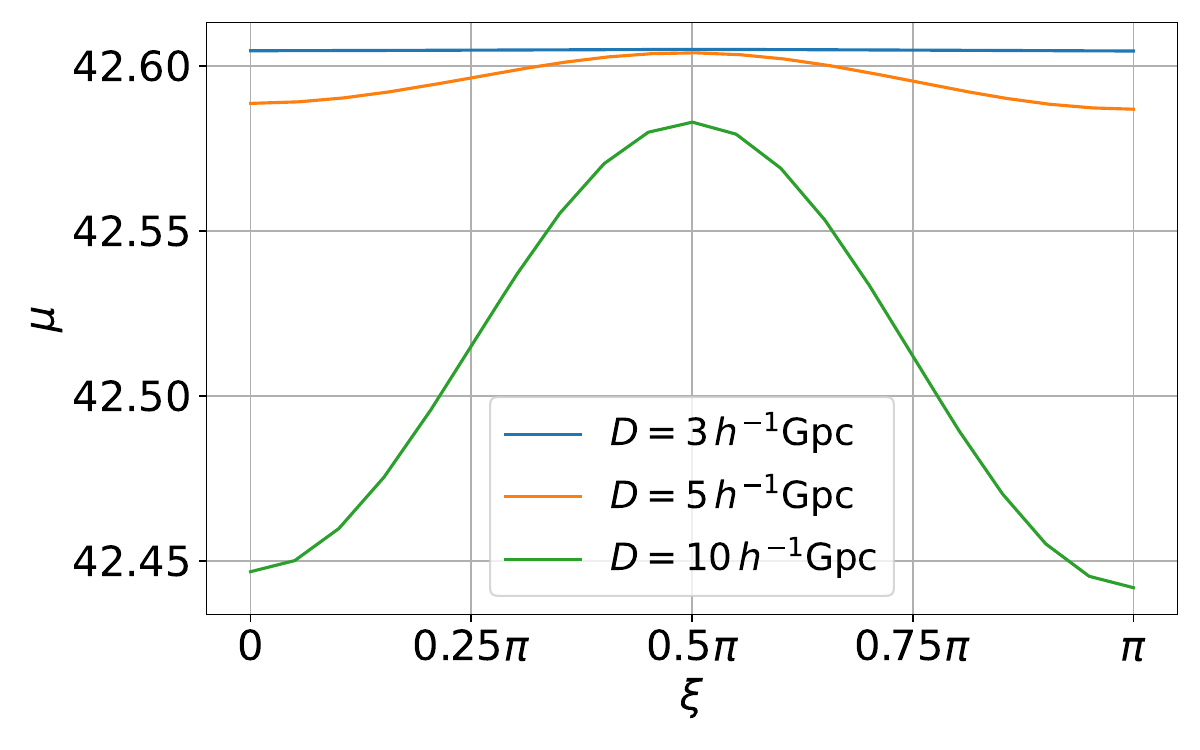}
    \end{minipage}
    \begin{minipage}{0.45\columnwidth}
        \centering
        \includegraphics[width=0.95\linewidth]{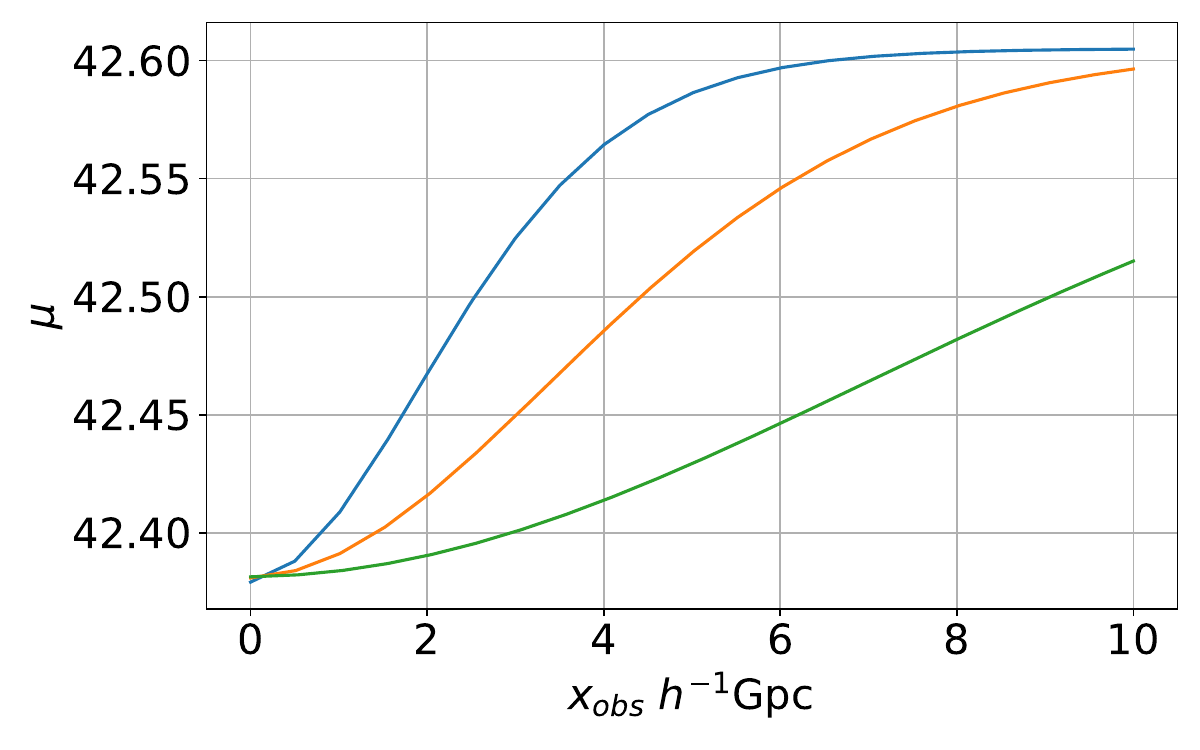}
    \end{minipage}
    \end{tabular}
\caption{Distance modulus $\mu$ for $\tilde{\Omega}_{\DW}=0.2$, $\mathsf{z}=0.5$, and $h=0.7$. The inclination $\xi$ varies with $x_{\rm obs} = 10h^{-1}\,\si{Gpc}$
fixed in the left panel, while $x_\obs$ varies with $\xi=\pi/4$ fixed in the right panel.
For $D \lesssim x_{\rm obs}$, the result is nearly $\Lambda$CDM with negligible angular dependence. For $D \gtrsim x_{\rm obs}$, the DW induces angular variations.  
The colour coding is the same in both panels.}
\label{mu}
\end{figure}

Figure~\ref{mu} shows the distance modulus $\mu$ for $\tilde{\Omega}_{\DW}=0.2$, $\mathsf{z}=0.5$, and $h=0.7$ as a function of the thickness $D$ of the DW, the angle $\xi$, and the observer's distance $x_\obs$.
In the left panel, when the thickness of the DW is smaller than $x_{\rm obs}$, the result is nearly indistinguishable from $\Lambda$CDM (as seen in figure~\ref{geodesic}), so the angular dependence is negligible. However, when $D$ exceeds $x_{\rm obs}$, the DW still exerts a noticeable influence, producing a significant angular dependence.  
In the right panel, as $x_{\obs}$ becomes sufficiently larger than the thickness of the DW, the model approaches the limit of $\Lambda$CDM, and the distance modulus $\mu$ shows
little variation.

\begin{figure}
    \centering
    \begin{tabular}{c}
        \begin{minipage}{0.95\hsize}
            \centering
            \includegraphics[width=0.95\linewidth]{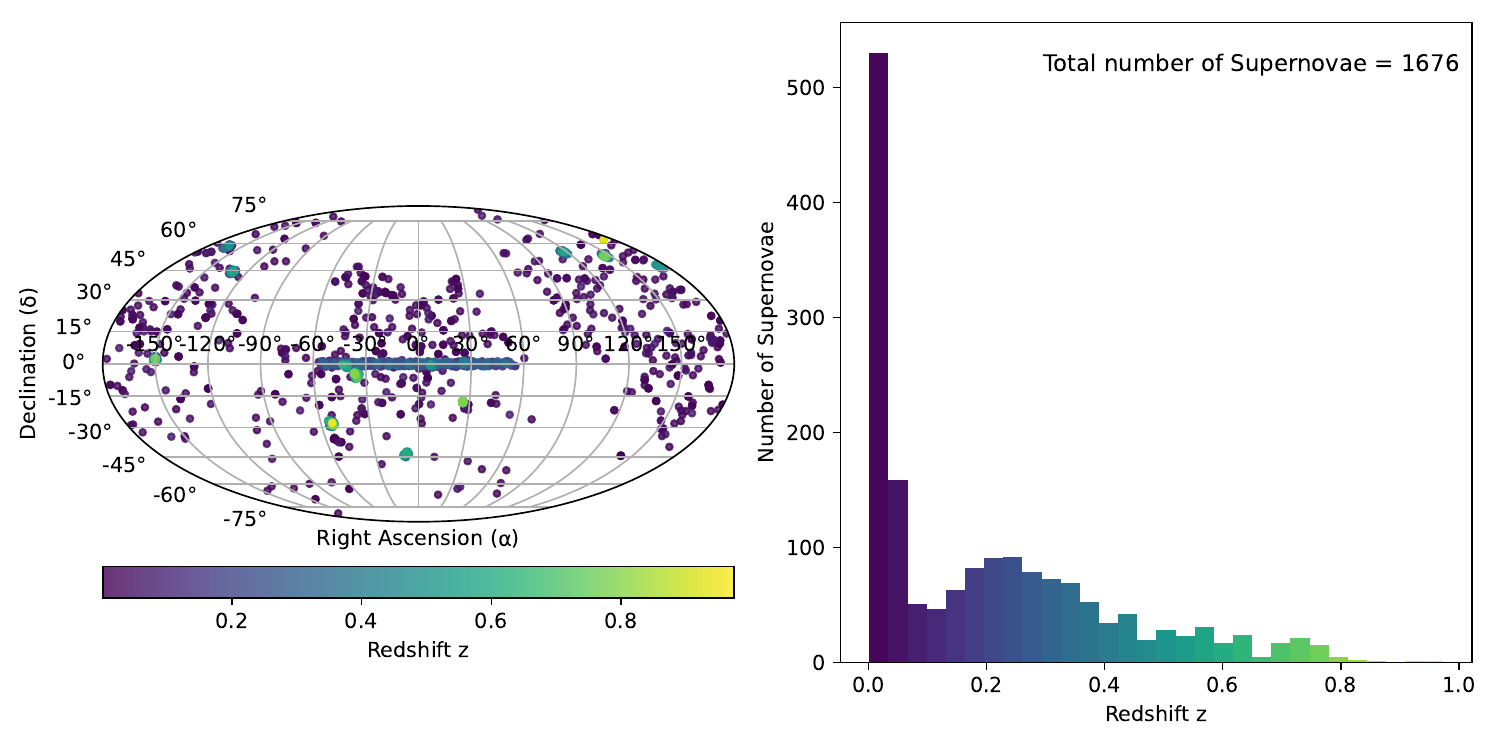}
        \end{minipage} \\
        \begin{minipage}{0.95\hsize}
            \centering
            \includegraphics[width=0.95\linewidth]{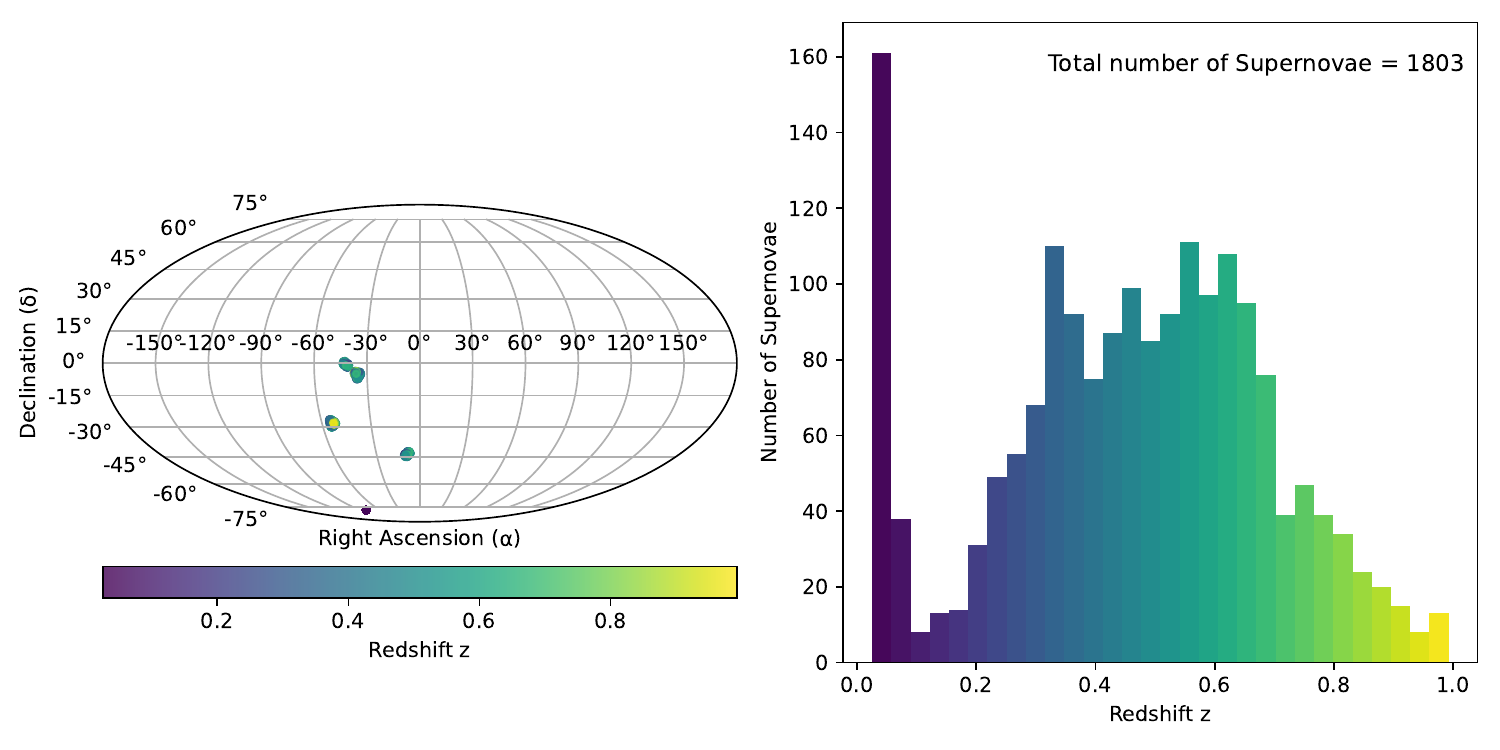}
        \end{minipage}
    \end{tabular}
\caption{\emph{Top}: Pantheon+~SH0ES ($z < 1$) data sets. \emph{Bottom}: DESY5 data. \emph{Left}: positions and redshifts of SNe Ia ($z < 1$), projected onto equatorial coordinates, with colour indicating redshift. \emph{Right}: histograms showing the number of SNe Ia in each redshift bin.}
\label{SNe}
\end{figure}

Figure~\ref{SNe} shows the locations and redshifts of SNe Ia used in \ac{MCMC} analysis. We utilised these observational data to estimate key model parameters such as the thickness $D$ of the DW, the energy density parameter $\tilde{\Omega}_{\DW}$, the Hubble parameter $h$, and the coordinates $(\alpha_\cc, \delta_\cc)$ of the normal direction of the DW, along with the observer's distance $x_\obs$ from the DW.
To infer these parameters, we performed an \ac{MCMC} analysis. 
In order to ensure consistency with the CMB constraints, we select prior distributions that satisfy the CMB constraints, as shown in figure~\ref{multipole} and summarised in table~\ref{prior}.

\begin{table}
    \centering
    \caption{Prior distributions for \ac{MCMC} analysis.}
    \label{prior}
    \begin{tabular}{
    l c}
    \toprule
    Parameter & Prior Range \\
    \hline
    $D$ (DW thickness) & [3,~10.2] $h^{-1}\,\si{Gpc}$ \\
    $\tilde{\Omega}_{\DW}$ (DW energy density fraction) & [$10^{-3}$,~0.2] \\
    $h$ (Hubble parameter) & [0.4,~0.95] \\
    $x_{\rm obs}$ (Observer distance from DW) & [0,~10] $h^{-1}\,\si{Gpc}$ \\
    $\alpha_c$ (Right ascension of DW normal) & [0,~360] degrees \\
    $\delta_c$ (Declination of DW normal) & [-90,~90] degrees \\
    \bottomrule
    \end{tabular}
\end{table}

Figure~\ref{SNe} also provides a visualisation of the distribution of SNe Ia across different redshifts, including data from both the Pantheon+~SH0ES surveys (top panel), as well as DESY5 (bottom panel). The left-hand side of the figure shows the positions of the SNe Ia in equatorial coordinates, with a colour code representing their redshift values. The histograms summarise the number of SNe Ia in each redshift bin, providing an overview of the redshift distribution across the two observational datasets.

\begin{figure}
\centering
\includegraphics[width=0.8\linewidth]{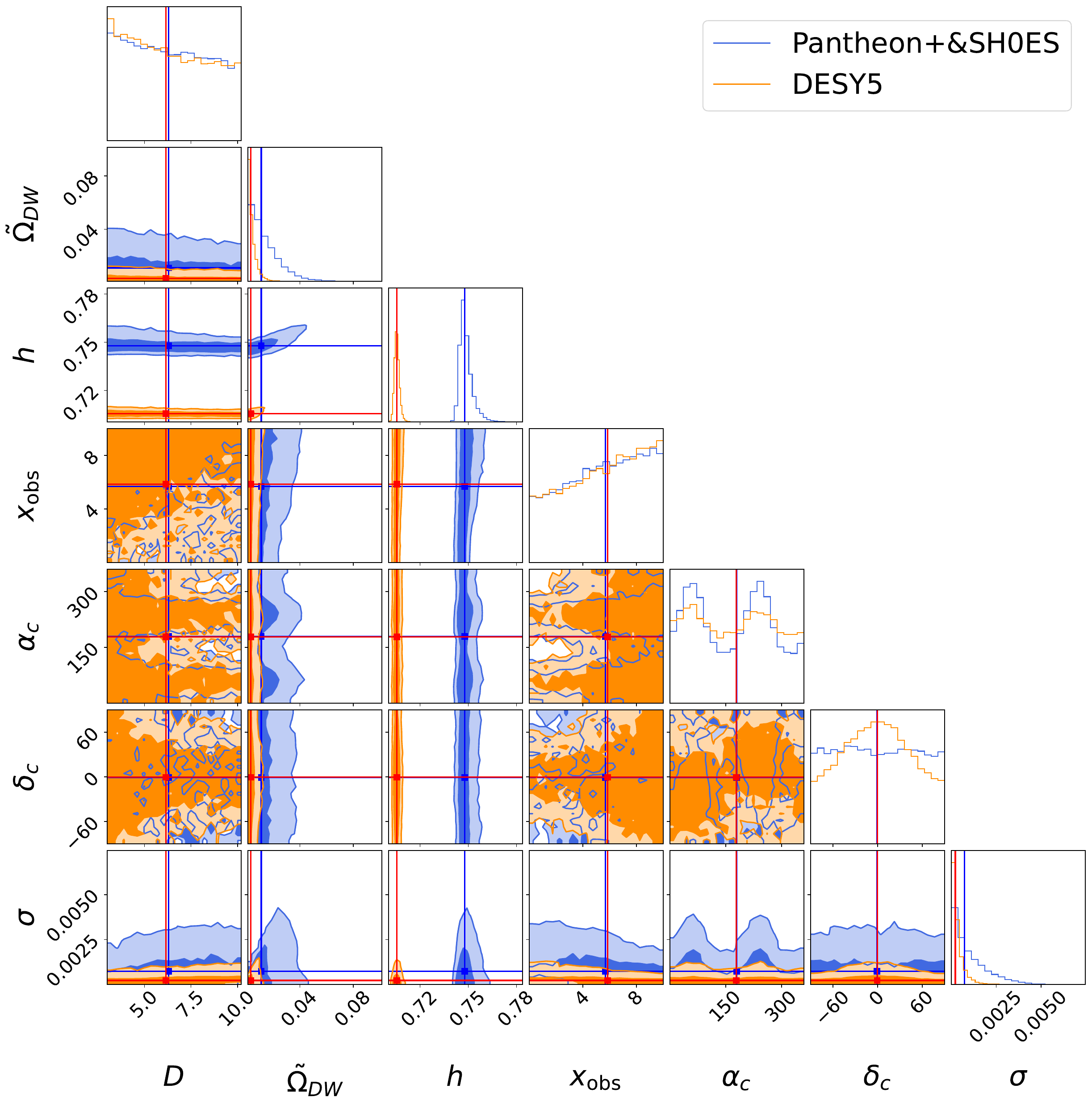}
\caption{Corner plot of the posterior distributions from the \ac{MCMC} analysis for the Pantheon+~SH0ES (blue) and DESY5 (orange) datasets. Each panel shows the one- and two-dimensional marginalised posterior distributions in the six-dimensional parameter space. The vertical and horizontal solid lines indicate the posterior medians (50th percentiles) of the parameters for each dataset. The contours enclose regions containing 68\% and 95\% of the posterior probability. The parameter $\sigma$ represents the anisotropy of the expansion at the observer's position, derived from the sampled parameters.}
\label{MCMC_pantheonDES}
\end{figure}

The results of the \ac{MCMC} analysis are shown in figure~\ref{MCMC_pantheonDES}, which presents the marginalised posterior distributions for the model parameters obtained from both the ``Pantheon+~SH0ES'' and ``DESY5'' datasets.
The vertical and horizontal lines represent the median (50th percentile) of each parameter's posterior distribution, while the shaded contours indicate the 68\% and 95\% credible regions that enclose the corresponding fractions of the total posterior probability in the two-dimensional parameter subspaces.
For the joint analysis of both datasets, see appendix~\ref{jointed}.

Additionally, we employ the Akaike Information Criterion (AIC) and the Bayesian Information Criterion (BIC) to compare our model with the standard $\Lambda$CDM model. The results, shown in table~\ref{likelihood}, demonstrate that the $\Lambda$CDM model is favoured not only by the CMB data but also by the SNe Ia data. Both AIC and BIC values indicate a preference for the $\Lambda$CDM model over the planar DW model, with positive differences in AIC and BIC for the DW model in both datasets.

\begin{table}
    \centering
    \caption{AIC and BIC values for the Pantheon and DES data sets.}
    \label{likelihood}
    \begin{tabular}{l c c c c}
    \toprule
    & \multicolumn{2}{c}{Pantheon} & \multicolumn{2}{c}{DES} \\
    Model & AIC & BIC & AIC & BIC \\
    \hline
    $\Lambda$CDM & 738 & 748 & 3422 & 3433 \\
    Planar DW & 750 & 782 & 3452 & 3485 \\
    Difference & +12 & +34 & +30 & +52 \\
    \bottomrule
    \end{tabular}
\end{table}

\section{Conclusions and discussion}\label{conclusion}

In this work, we discuss a Hubble-scale planar domain-wall DE model and examine its dynamical and observational consequences.
We have shown that, near the central plane of the domain wall, the spacetime is approximately homogeneous but intrinsically anisotropic, characterised by different expansion rates parallel and perpendicular to the wall.
This anisotropy provides a concrete physical origin for direction-dependent cosmic acceleration.

We confronted the model with current observational data.
The anisotropic expansion inevitably induces CMB temperature anisotropies, and constraints from the Planck 2018 dipole, quadrupole, and octopole measurements place stringent upper bounds on the DW energy density.
Only a parameter region in which the DW contribution is extremely small and the observer is located far from the wall is consistent with the observed CMB sky.

We further performed a Markov Chain Monte Carlo analysis using SNe Ia data, including the Pantheon+~SH0ES and DESY5 samples, within the parameter region allowed by the \ac{CMB} constraints.
The results favour the $\Lambda$CDM model, in which the DW abundance is negligible, and the universe is isotropic.

Our results imply that, while a Hubble-scale planar DW provide an interesting and physically motivated realisation of anisotropic DE, their contribution to the present cosmic acceleration must be highly suppressed.
Future observations with improved sensitivity to large-scale anisotropies may further test such scenarios, but within current data, $\Lambda$CDM remains the preferred description of the late-time universe.

Several extensions of this work are worth pursuing.
It would be interesting to study the impact of DW-induced anisotropic expansion on CMB polarisation and higher-order statistics, which may provide probes more sensitive than temperature anisotropies alone.
Correlations between the CMB and large-scale structure, such as the
integrated Sachs--Wolfe effect, also offer a promising avenue to further constrain or detect residual anisotropic signatures.

On the theoretical side, generalisations to moving or slightly curved DWs, as well as a more explicit modelling of their formation mechanism, could clarify the robustness of the present constraints.
Finally, a systematic comparison with other anisotropic DE models, together with future supernova and standard-siren observations, would help to establish a comprehensive observational framework for testing anisotropic cosmic acceleration.

\acknowledgments 
NK thanks Shun Yoshioka and Shintaro Hayashi for their insightful discussions on the theoretical aspects of this work. NK is also grateful to Keitaro Ishikawa for his guidance on the \ac{MCMC} analysis. Furthermore, NK appreciates the valuable advice and useful discussions provided by the members of the Nagoya University Cosmology Group, particularly through questions and feedback during seminars.
This work was supported by JSPS KAKENHI JP24K07047 (YT), 25H02165
(FT), 26K00695 (FT), and 26H00403 (TT). TT's work was partly supported by the 34th (FY 2024) Academic research grant (Natural Science) No.~9284 from DAIKO FOUNDATION.
This work was also supported by the World Premier International Research Center Initiative (WPI), MEXT, Japan, and is based upon work from COST Action COSMIC WISPers CA21106, supported by COST (European Cooperation in Science and Technology).

\appendix

\section{Approximate solutions for near and far DW region}\label{near}

In Subsection~\ref{numerical}, we solve the unstable partial differential equations. In this section, we show that the numerical results are consistent with the analytical solutions, demonstrating the reliability of the numerical approach.

We now consider sub-leading corrections for the near-\ac{DW} solution with respect to the distance $x$.
We expand the relevant quantities around $x=0$ up to $\mathcal{O}(x^2)$ as follows.
\begin{align}
\Phi(t,x) &= c(t)\,x, \\
A(t,x) &= a_0(t) + a_2(t)\,x^2, \\
B(t,x) &= b_0(t) + b_2(t)\,x^2, \\
\rho_\m(t,x) &= \rho_0(t) + \rho_2(t)\,x^2 .
\end{align}
From eqs.~(\ref{abar}) and (\ref{newhubble}), the conservation equation (\ref{conserve2}) yields
\begin{align}
  \rho_0(t) = \rho_0(t_\ui)\,\bar{a}^{-3}.
\end{align}
Using these relations, the equations of motion at $\mathcal{O}(x^2)$ reduce to
\begin{align}
  \frac{8\pi}{\mpl^2}
  \left(
  \frac{\lambda\eta^4}{4}
  + \rho_0(t_\ui)\bar{a}^{-3}
  + \rho_\Lambda
  \right)
  &= 3H_{\bar{a}}^2,
  \label{hamapp} \\
  \frac{\pi}{\mpl^2}
  \left(\lambda\eta^4 + 4\rho_\Lambda\right)
  - \frac{2\pi c^2}{\mpl^2 a_0^2}
  &=
  \frac{1}{2}\left(\frac{\dot{b}_0}{b_0}\right)^2
  + \frac{\ddot{b}_0}{b_0},
  \label{oddapp} \\
  \ddot{c}
  + 3H_{\bar{a}}\,\dot{c}
  - \lambda\eta^2 c
  &= 0.
  \label{ELapp}
\end{align}
Equations~(\ref{hamapp}) and (\ref{ELapp}) can be solved with the initial conditions,
\begin{align}
  \bar{a}(t_\ui)=1, \qquad c(t_\ui)=\frac{\sqrt{\lambda}\eta^2}{\sqrt{2}}, \qquad
  \dot{c}(t_\ui)=0,
\end{align}
which correspond to an initially flat, homogeneous, and isotropic universe with a static domain wall.

We see that the relative error of the approximate solution compared to the numerical solution,
\begin{align}
 \frac{c(t)x-\Phi(t,x)}{\Phi(t,x)},
\end{align} 
is less than $\mathcal{O}(10)$\% for $x\ll D$.
To obtain the scale factors, we introduce the shear parameter
\begin{align}
  \varsigma_0 \equiv \frac{b_0}{a_0}.
\end{align}
Substituting this definition into eq.~(\ref{oddapp}) yields a second-order differential equation for $\varsigma_0$.
After solving it numerically, the scale factors are approximately obtained as
\begin{align}
  a_0 = \left(\frac{\bar{a}^3}{\varsigma_0^2}\right)^{1/3},
  \qquad
  b_0 = \left(\bar{a}^3\varsigma_0\right)^{1/3}.
\end{align}
We find that the error of the approximate solution remains $\mathcal{O}(1)\%$ throughout the evolution.

\medskip
Next, we consider an approximate solution in a region far away from the DW.
In the asymptotic regime ($|\bar x|\gg 1$), the scalar field approaches its vacuum values,
\begin{align}
  \Phi(\pm\infty,t)=\pm\eta,
\end{align}
and the influence of the DW becomes negligible.

We define the perturbation $\delta\Phi \equiv \Phi-\eta$ and neglect spatial derivatives.
Then, from eqs.~(\ref{mom}) and (\ref{EL}), we obtain the following solutions:
\begin{align}
  v&=0,  \\
  \delta\Phi
  &= C_1\cos(\sqrt{2\lambda}\eta\,t)
  + C_2\sin(\sqrt{2\lambda}\eta\,t),
  \label{deltaPhi}
\end{align}
where $C_1$ and $C_2$ are integration constants determined by the initial conditions.

Substituting eq.~(\ref{deltaPhi}) into eq.~(\ref{odd}), we obtain
\begin{align}
  \dot{K}_2{}^2-\frac{3}{2}(K_2{}^2)^2
  =
  \frac{4\pi\lambda\eta^2}{\mpl^2}
  \left[
  (-C_1^2+C_2^2)\cos(2\sqrt{2\lambda}\eta\,t)
  -2C_1C_2\sin(2\sqrt{2\lambda}\eta\,t)
  -\frac{\rho_\Lambda}{\lambda\eta^2}
  \right].
  \label{K22}
\end{align}
During the accelerated expansion phase, the following condition is satisfied.
\begin{align}
  2\sqrt{2\lambda}\eta\,t \ll 1
\end{align}
Expanding the trigonometric functions to the leading order, eq.~(\ref{K22}) reduces to
\begin{align}
  \dot{K}_2{}^2-\frac{3}{2}(K_2{}^2)^2
  = \chi-\psi t,
  \label{K22approx}
\end{align}
where
\begin{align}
  \chi &\equiv
  \frac{4\pi\lambda\eta^2}{\mpl^2}
  \left(-C_1^2+C_2^2-\frac{\rho_\Lambda}{\lambda\eta^2}\right),
  \label{chi}\\
  \psi &\equiv\frac{16\sqrt{2}\pi\lambda^{3/2}\eta^3}{\,\mpl^2}\,C_1C_2.
  \label{psi}
\end{align}
The quantity $K_2{}^2$ can be obtained by solving eq.~(\ref{K22approx}).
On the other hand, using eqs.~(\ref{ham}) and (\ref{sum}), we find
\begin{align}
  2\dot{K}-2K^2+6K_2{}^2K-9(K_2{}^2)^2
  \simeq
  \frac{8\pi\lambda\eta^2}{\mpl^2}
  \left[
  4\sqrt{2\lambda}\eta\,C_1C_2\,t
  + C_1^2-2C_2^2
  -\frac{3\rho_\Lambda}{\lambda\eta^2}
  \right].
  \label{K}
\end{align}
The function $K$ can then be determined by solving eq.~(\ref{K}).

As initial conditions, we assume
\begin{align}
  \Phi(x,t_\ui)
  &=
  \eta\tanh(\frac{\sqrt{\lambda}\eta\,x}{\sqrt{2}}),
  \qquad
  \dot{\Phi}(x,t_\ui)=0,
\end{align}
and introduce a small parameter $\epsilon$ such that $\delta\Phi(x,t_\ui)=\epsilon$.
This leads to
\begin{align}
  C_1 &= \epsilon\cos(\sqrt{2\lambda}\eta\,t_\ui),\qquad
  C_2 = \epsilon\sin(\sqrt{2\lambda}\eta\,t_\ui).
\end{align}
With these expressions, we obtain 
\begin{align}
  \chi &=
  -\frac{4\pi\lambda\eta^2}{\mpl^2}
  \cos(2\sqrt{2\lambda}\eta\,t_\ui)\,\epsilon^2
  -\frac{4\pi}{\mpl^2}\rho_\Lambda, \\
  \psi &=
  \frac{8\sqrt{2}\pi\lambda^{3/2}\eta^3}{\mpl^2}
  \sin(2\sqrt{2\lambda}\eta\,t_\ui)\,\epsilon^2
\end{align}
Since $\epsilon$ is small, terms of order $\mathcal{O}(\epsilon^2)$ can be neglected at the leading order.
Using the relation $K_2{}^2=-\dot{B}/B$, eq.~(\ref{K22approx}) simplifies to
\begin{align}
  \frac{\ddot{B}}{B}
  =
  -\frac{1}{2}\left(\frac{\dot{B}}{B}\right)^2
  +\frac{\Lambda}{2}.
  \label{K22ana}
\end{align}
This equation is identical to the acceleration equation in the $\Lambda$CDM model. Substituting this into eq.~(\ref{K}) yields the same expression for the scale factor $A$.
Therefore, we conclude that the universe asymptotically behaves as a $\Lambda$CDM universe far away from the DW.

\section{CMB higher multipoles}
\label{higher}

\begin{figure}
\centering
\includegraphics[width=0.95\linewidth]{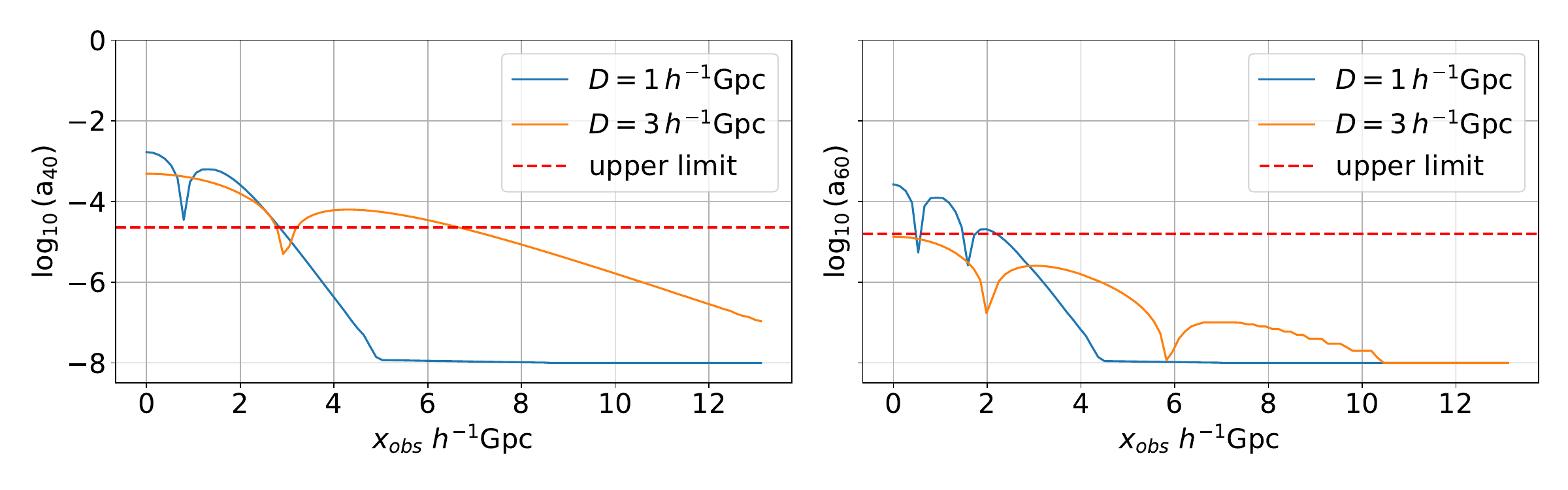}
\caption{Dependence of the coefficients $a_{40}$ (left) and $a_{60}$ (right) on the observer position $x_{\mathrm{obs}}$ for $D=1h^{-1}\,\si{Gpc}$ (blue) and $3h^{-1}\,\si{Gpc}$ (orange), assuming $\tilde{\Omega}_{\DW}=0.05$. The red dashed line shows the upper bound of the Planck 2018 observation results.
The coefficient $a_{40}$ exhibits a local minimum around $x_{\mathrm{obs}}\approx D$, while $a_{60}$ shows a local maximum near this position, with local minima on both sides.
}
\label{higher_multi}
\end{figure}

In this section, we discuss the higher multipoles of the \ac{CMB} anisotropies.  
From figure~\ref{multipole},
we see that even multipoles constrain the model parameters more stringently than odd multipoles.  
Figure~\ref{higher_multi} shows the dependence of the coefficients hexadecapole ($a_{40}$) and triakontadipole ($a_{60})$ on the position of the observer $x_{\mathrm{obs}}$ for several values of the parameter $D$, assuming $\tilde{\Omega}_{\DW}=0.05$. 

\begin{figure}
\centering
\includegraphics[width=0.8\linewidth]{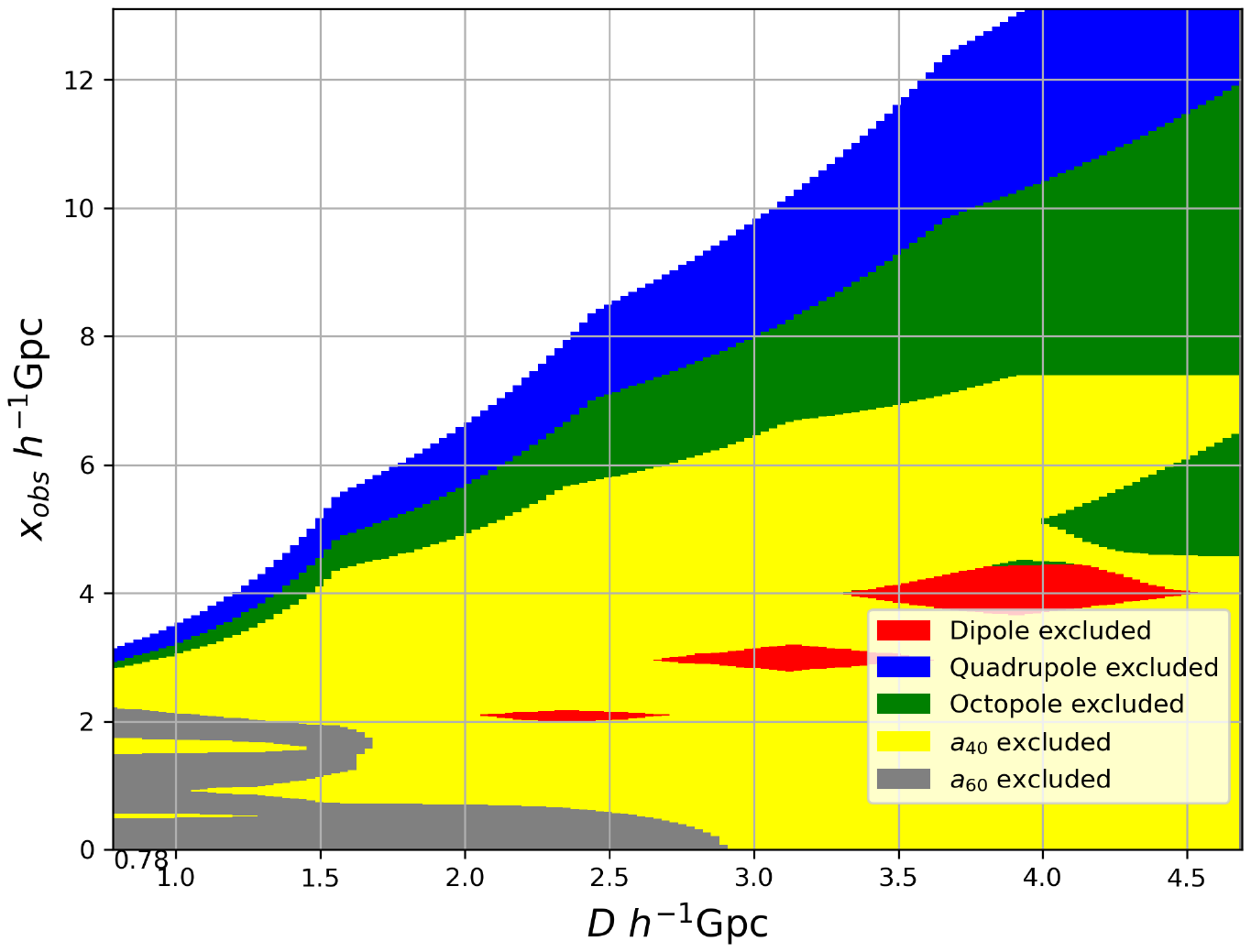}
\caption{
Excluded parameter regions from the CMB dipole (red), quadrupole (blue), octopole (green), hexadecapole ($a_{40}$; yellow), and triakontadipole ($a_{60}$; grey) constraints for $\tilde{\Omega}_{\DW}=0.05$. The dipole region is connected behind the $a_{40}$ region.
The quadrupole provides the most stringent constraint on the model parameters, favouring the isotropic universe.
}
\label{thin}
\end{figure}

We constrain the model parameters using these higher multipoles.  
The method is the same as that described in section~\ref{CMB}.
\begin{align}
    a_{40}^{\mathrm{obs}}
    &=
    \sqrt{
    \frac{9\,C_{4,\mathrm{obs}}}
    {(2.7255\times 10^{6}~\si{\mu K})^2}
    }
    \le 2.27\times10^{-5},\\
    a_{60}^{\mathrm{obs}}
    &=
    \sqrt{
    \frac{13\,C_{6,\mathrm{obs}}}
    {(2.7255\times 10^{6}~\si{\mu K})^2}
    }
    \le 1.56\times10^{-5}.
\end{align}
Figure~\ref{thin} shows the constraints from the CMB anisotropy for $\tilde{\Omega}_{\DW}=0.05$.  
The shaded region does not satisfy these constraints.  
Compared to figure~\ref{multipole} (left), the allowed region does not change.  
We find that the quadrupole constraint is the most stringent, so we show only the constraints up to the quadrupole in the main text.

\section{MCMC with joint data}\label{jointed}

\begin{figure}
\centering
\includegraphics[width=0.8\linewidth]{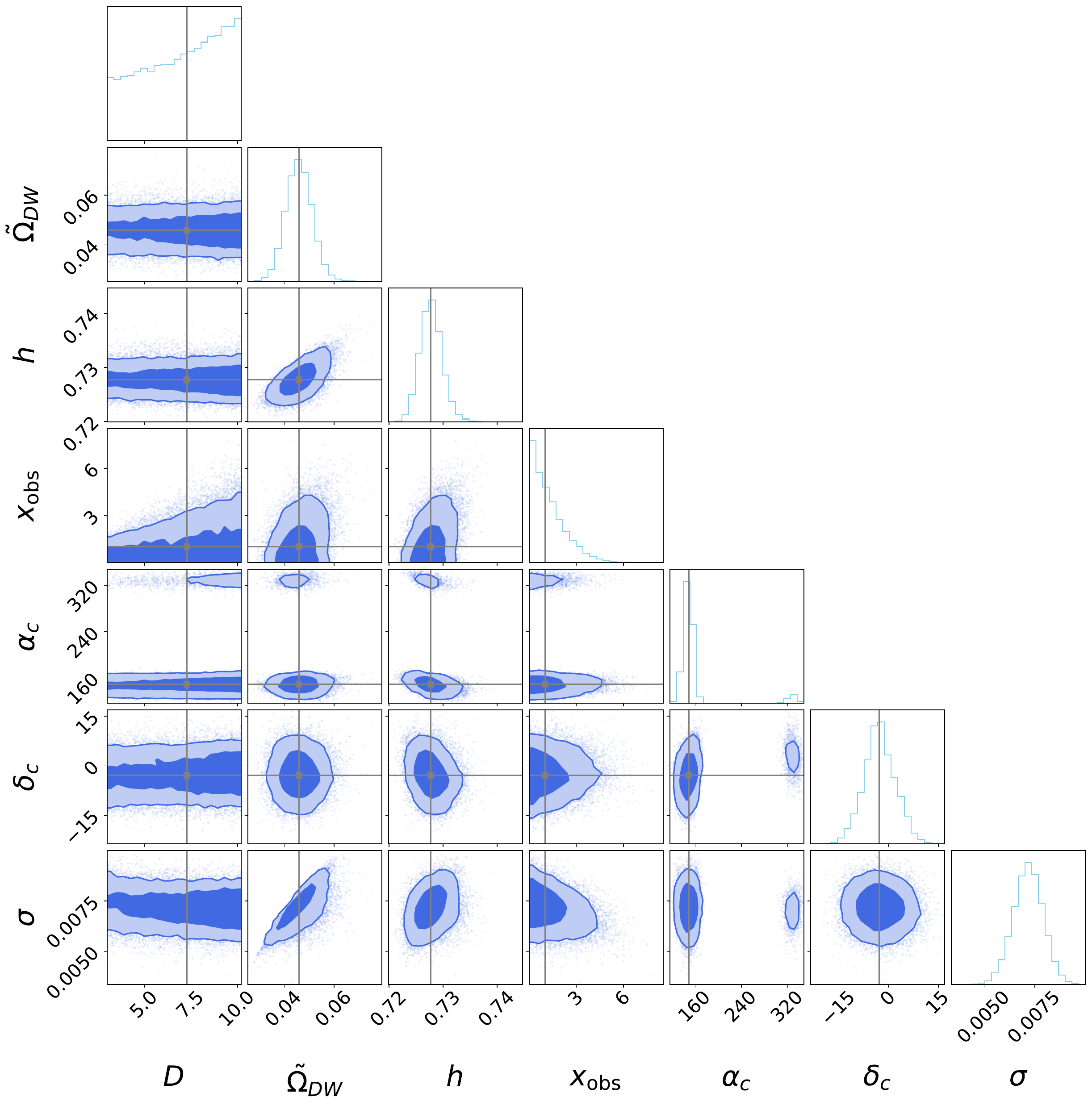}
\caption{Posterior distributions from the joint \ac{MCMC} analysis (see~\ref{MCMC}) using the same priors as in table \ref{prior}. The SNe Ia sample is the combined dataset shown in figure~\ref{SNe}, excluding 378 overlapping SNe. Nonzero $\tilde{\Omega}_{\DW}$ and $\sigma$ suggest anisotropy; however, this arises from differences in the inferred $h$ between the two observations (see figure~\ref{MCMC_pantheonDES}) and should be interpreted with caution compared to the individual \ac{MCMC} results.}
\label{MCMC_both}
\end{figure}

Figure~\ref{MCMC_both} presents the results of the joint \ac{MCMC} parameter estimation using the ``Pantheon+~SH0ES" and ``DESY5" datasets (see section~\ref{MCMC}). The priors adopted in this analysis are identical to those listed in table~\ref{prior}. The SNe Ia sample is constructed by combining the datasets shown in figure~\ref{SNe}, excluding 378 overlapping SNe. Figure~\ref{MCMC_both} displays the resulting posterior distributions. Notably, the parameters $\tilde{\Omega}_{\DW}$ and $\sigma$ are inferred to be nonzero, which potentially suggests an anisotropic universe favoured by the SNe observations. However, this apparent signal could arise from the difference in the inferred values of the Hubble parameter $h$ between the two observational datasets (see figure~\ref{MCMC_pantheonDES}), which show deviations of the angular directions of the sources.
We note that a robust assessment of anisotropy using a joint analysis would, in principle, require a prior cross-calibration of the two observational datasets before combining them, as otherwise residual systematic differences could be misinterpreted as anisotropic signatures.

\bibliographystyle{JHEP} 
\bibliography{material,DE,align,anisotropySNe}

\end{document}